\documentclass[article,draftcls,onecolumn]{IEEEtran}
\usepackage{amsmath,amssymb}
\usepackage{graphicx,color}
\usepackage{subfig}
\usepackage{cite}
\usepackage{epstopdf}
\usepackage{url,doi}
\usepackage{caption}
\usepackage[colorinlistoftodos,color=orange!70,textsize=scriptsize,shadow]{todonotes}
\usepackage{setspace}

\usepackage[normalem]{ulem}
\newcommand\redout{\bgroup\markoverwith{\textcolor{red}{\rule[.5ex]{2pt}{0.4pt}}}\ULon}

\newtheorem{theorem}{Theorem}

\newtheorem{proposition}{Proposition}

\newtheorem{assumption}{\textbf{Modelling Assumption}}

\begin{document}

\title{Market Segmentation for Privacy Differentiated ``Free" Services}
\author{
Chong Huang, 
\and Lalitha Sankar 
\thanks{C. Huang and L. Sankar are with the Department of Electrical, Computer, and Energy Engineering at Arizona State University, Tempe, AZ 85287 (e-mail: \texttt{chong.huang@asu.edu}, \texttt{lalithasankar@asu.edu}). }%
}

\maketitle

\pagenumbering{arabic}

\begin{abstract}
The emerging marketplace for online free services in which service providers earn revenue from using consumer data in direct and indirect ways has lead to significant privacy concerns. This leads to the following question: can the online marketplace sustain multiple service providers (SPs) that offer privacy-differentiated free services? This paper studies the problem of market segmentation for the free online services market by augmenting the classical Hotelling model for market segmentation analysis to include the fact that for the free services market, a consumer values service not in monetized terms but by its quality of service (QoS) and that the differentiator of services is not product price but the privacy risk advertised by a SP. Building upon the Hotelling model, this paper presents a parametrized model for SP profit and consumer valuation of service for both the two- and multi-SP problems to show that: (i) when consumers place a high value on privacy, it leads to a lower use of private data by SPs (i.e., their advertised privacy risk reduces), and thus, SPs compete on the QoS; (ii) SPs that are capable of differentiating on services that do not directly target consumers gain larger market share; and (iii) a higher valuation of privacy by consumers forces SPs with smaller untargeted revenue to offer lower privacy risk to attract more consumers. The work also illustrates the market segmentation problem for more than two SPs and highlights the instability of such markets.

\textbf{Keywords-}free online services, privacy-differentiated services, quality of service, market segmentation.
\end{abstract}
\section{Introduction}
There has been a steady increase in online interactions between consumers and retailers, where the term retailers refers to entities who sell or offer (for free) a product. In fact, many oft used online services are offered for free and consumers implicitly accede to tracking for customized services. Targeted ads are a part of the emerging revenue/profit model for such retailers, henceforth referred to as \textit{service providers} (SPs), especially those offering free services. 
Consumers are delighted by free services until they begin encountering privacy violations on a daily/frequent basis. While such infractions taken individually could be ignored or discounted, the totality of data available about consumers with a variety of retailers and the resulting privacy consequences raise serious concerns. 


Service providers are beginning to acknowledge that consumers are sensitive to privacy violations. For example, Google \cite{Erlingsson2014} and Apple \cite{AppleSupport2016} recently adopted differentially private mechanisms for collecting user data for statistical analyses. However, the details of these mechanisms are opaque and offer even less clarity on whether the consumer actually has a choice. In this context, it is worth understanding if privacy-differentiated services can provide such choices for consumers. In a competitive marketplace, the aggregate weight of targeting may drive some customers to seek a more privacy-protective alternative. The cost to the consumer for this action may be a lower quality of service (QoS) (e.g., poorer search engine capabilities). However, it could eventually lead to a more open model for consumer sharing of private information, i.e., one from implicit assent to informed consent. 

In this paper, we identify and formalize conditions under which privacy-differentiated services are sustainable and examine when \textit{competition} based on privacy protections could lead to a sustainable marketplace for online free services. As a hypothetical running example we posit a marketplace with two search engines Google Search and DuckDuckGo. The former tracks searches by users thereby offering higher quality service (QoS) (e.g., search accuracy) while the latter explicitly does no tracking and may offer a relatively lower QoS. More generally, our model allows differentiating SPs by their QoS and privacy-sensitive offerings to determine the existence of a stable market for privacy-differentiated services. We apply and build upon classical game theoretic methods, in particular the Hotelling model for market segmentation, to quantify the market segmentation. 
We also generalize the model to multiple SPs (e.g., Google, DuckDuckGo, and Bing) and illustrate the instability of multi-competitor markets.

\subsection{Related Work}
An extensive body of literature on economic models for privacy was recently reviewed by Acquisti \textit{et al.}~\cite{AcquisiTW:16econ}. These models illustrate the large semantic range covered by the word ``privacy".  Targeted advertising is a common method for service providers to exploit knowledge of a consumer in a way that can cause privacy violations. 
Our work is informed by the literature on targeting strategies for retailers~\cite{shaffer1995competitive,chen2002research,tang2008gaining,campbell2015privacy,conitzer2012hide,chen2001individual,jentzsch2012study,lee2011managing,chellappa2010mechanism,DattaTD:pets2015}, but rather than optimizing retailer strategies we are interested in identifying how privacy-differentiated services can address privacy concerns. 

The problem of market segmentation is a classic and well-studied problem in microeconomics \cite{perloff2016microeconomics} with focus on how pricing and product differentiation can lead to a stable and competitive marketplace. However, the free online services market present a new challenge wherein monetary quantification of both `free' services and the data collected about consumers is not simple and straightforward. Equally challenging is the quantification of consumer privacy since it requires capturing the heterogeneous expressions of privacy sensitivity that can range from `don't care' at one extreme to `hyper vigilant' at the other. However, some aspects of market models can be brought to bear to our problem; in particular, the oligopolistic market model with a small number of competitors, barriers to entry that are not as high as those for monopolies, and with differentiated products fits appropriately for the markets we are considering wherein two or (a few) more service providers offer products of the same type but differentiated by QoS and privacy risk.


For a two-player oligopolistic market game, the Cournot-Nash and Bertrand duopoly models are considered classic models wherein the two firms differentiate using quality and price, respectively. 
A more nuanced model that captures differentiation between two firms and consumer preferences is the Hotelling model \cite{hotelling1990stability}. This model captures differentiation between market players by mapping firms to positions on a unit length line such that the location is indicative of the firm's `differentiation level', the total line length is reflective of the entire market, a consumer's privacy preference is a point on the line, and the optimal locations of the firm resulting from the simultaneous game between the players indicate the resulting segmentation. The model captures utility for consumer as both the advantage (price, quality, etc) from the firm as well as the `distance cost' from the consumer's location to that firm. Consumers choose the seller which give them the highest utility (in terms of their valuation of the product and its price as well as the `transportation' costs). The Hotelling model has also been extended to include gradations in product (quality) and customer types via a vertical variant of the model \cite{wauthy1996quality} . 

\textbf{Privacy and market segmentation.} Jentzsch et.al.~\cite{jentzsch2012study} propose a model to study competitions between two service providers by taking consumer's privacy preference (binary choices: low privacy/high privacy) into account using a vertical Hotelling model. Thus, consumers select the service provider based on their privacy concerns and the amount of payment to the service provider. They provide analysis of equilibrium strategies for SPs. In \cite{lee2011managing}, Lee \textit{et al.} study the influence of privacy protection on the segmentation of a duopoly. In their model, firms may offer standard and personalized products with personalized prices to three different types of privacy-sensitive consumers (the `unconcerned' who always share information, `pragmatic' ones who only share if a firm adopts privacy protection, and the `fundamentalists' who never share data). They show that a privacy-friendly firm can enlarge market share by attracting more pragmatists to share personal information. From this expansion it can earn more profits rather than compete with its rival for the other consumers. In contrast to both above-mentioned models, our model differs in focusing on `free' services, and thus, introduces new models for quantifying QoS- and privacy-based differentiators; furthermore, our model generalizes the discrete set of privacy sensitive consumers in \cite{lee2011managing} to a continuous set of privacy risks thus allowing analysis of over an entire range of privacy expression and present a more nuanced view of how SPs should offer services to all types of consumers. 

\subsection{Our Contributions}
Our work introduces a game-theoretic interaction model for free online services offered by two or more SPs with the goal of understanding whether privacy-differentiated service offerings have the ability to capture market share. Our model captures a variety of free online services such as search engines, social networking sites, and software apps that are free, and therefore, use consumer data in a variety of ways for revenue generation. Specifically, our model is based on the `spatial' Hotelling model wherein the location is now proxy for both the privacy risk levels that the SPs offer and consumers prefer (both often at odds). The QoS of the service now models the classical product price. Our model differentiates itself from the Hotelling model in the following sense: unlike the classical model of non-negative transportation costs from consumer `location' (preference) to either SP `location', a consumer with a specific privacy risk choice gains from choosing an SP with a lower risk offering and loses from choosing one with a higher risk offering. This in turn leads to different outcomes than the classical model; we use a three-stage sequential game to compute the optimal strategies for the SPs and the resulting market share for specific models of cost and revenue (to SPs), distribution of consumer heterogeneous privacy choices, as well as QoS valuation (to consumers). 

We present closed form solutions for the two SP market with linear valuation functions (cost, revenue, consumer utility) and a uniform distribution of consumer preferences; for this settings, our results highlight the following: (i) when consumer place a high value on privacy, it leads to a lower use of private data by SPs, i.e., their advertised privacy risk reduces; (ii) SPs offering high privacy risk services are sustainable only if they offer sufficiently high QoS; (iii) SPs that are capable of differentiating on services that do not directly use consumer data gain larger market share; and (iv) higher consumer privacy valuation forces SPs with smaller privacy-independent (untargeted) revenue to offer lower privacy risk service to attract more consumers. In extending the work to more than two SPs, we illustrate the instability of such markets and highlight the challenges of studying market segmentation for more than two participants (a problem acknowledged in economics~\cite{brenner2005hotelling}). 
\subsection{Organization of the Paper}
The paper is organized as follows: Section~\ref{Section:SystemModel} introduces the system model and the non-cooperative game formulation. The main result for a two-SP market with linear valuation functions are presented in Section~\ref{Section:2SPs}. Section~\ref{Section:Multiple SPs} discuss equilibrium results for a market with multiple SPs. Finally, concluding remarks and future work are provided in Section~\ref{Section:Conclusions}. 
\section{Problem Model and Game Formulation}
\label{Section:SystemModel}
Our parametrized model detailed below captures the following privacy-differentiated market segmentation problem:
service providers offer free services differentiated by QoS and privacy risks. Online services that are offered for free often generate revenue by using the data they obtain from their consumers Their gain from using consumer data is captured by a revenue function and their cost of doing so is captured by a cost function. The goal of each SP is to choose a QoS and privacy risk tuple that maximizes its profit (difference of revenue and cost). The hetrogeneous expression of consumer privacy sensitivity is modeled as a (probability) distribution of the population over a range of privacy risk values. The consumer will choose the SP that maximizes a desired function of QoS, the privacy risk tolerance of the consumer, and the privacy risk offered by the SP. A stable strategy i.e., a strategy from which no participant will deviate without reduction in utility, of such a non-cooperative game will yield an optimal partition (market segmentation) of the consumers. We build upon the classical Hotelling model proposed by Hotelling in~\cite{hotelling1990stability} to study market segmentation.

Formally, we introduce a game-theoretic model for two SPs and infinitely many consumers. Each SP offers the same type of free services (e.g., search engine, social network) with a certain privacy risk guarantee $\varepsilon$ and quality of service (QoS) $v$. Thus, an SP differentiates its service by a tuple ($\varepsilon, v$) that it advertises to all consumers. The goal of this work is to determine the fraction of consumers (market segment) that choose each SP when the SPs offer ($\varepsilon, v$) tuples that maximize their profit. 

\subsection{Two-SP Market Model}

\subsubsection{SP Model}
\label{Section:SPS}
We consider two rational (i.e., profit maximization entities) SPs, denoted by $SP_1$ and $SP_2$. Both SPs provide the same kind of free service; but they differs in the QoS offered. Thus, $SP_1$ and $SP_2$ offer QoS $v_1$ and $v_2$, respective, where in general $v_1\neq v_2$. Furthermore, $SP_1$ and $SP_2$ guarantee that the privacy risk for using their services is at most $\varepsilon_1$ and $ \varepsilon_2$, respectively, where $\varepsilon_1,\varepsilon_2\in [0,\bar{\varepsilon}]$. Without loss of generality, we assume $\varepsilon_2\ge \varepsilon_1$. Under this consumption, $SP_2$ must offer a higher QoS ($v_2\ge v_1$). Otherwise, its strategy will be dominated by its opponent since $SP_1$ will offer both higher QoS and lower privacy risk. For example, $SP_1$ and $SP_2$ could be Duckduckgo and Google, respectively, in the search engine market, with the QoS given by the accuracy of search results. On the other hand, the privacy risk can correspond to different guarantees they provide on consumer data use; e.g., whether they will use consumer data only for statistical purposes or target consumers with tailored ads. We model this privacy risk guarantee as a variable taking values over a continuous range. In practice, such guarantees may be coarse granular choices; for example, between completely opting out of the targeting or allowing data use only for statistical purposes or complete data use only by SP or all possible data usage and sale. We assume that the SPs generate revenue in two ways: (i) by exploiting the private data of consumers to offer \textit{targeted} ads and other services to consumers; and (ii) by providing interested advertisers an online platform to reach consumers. This latter revenue is independent of private data and simply derived from the revenue capability of the platform.

Let $R_P(\varepsilon_i)$ denote the revenue of $SP_i$, $i\in\{1,2\}$, resulting from using the private data of consumers and let $R_{\text{NP},i}$ denote the revenue generated without using consumers' private information (e.g., from interested advertisers). The total revenue, $R(\varepsilon_i)$, of $SP_i$ from offering privacy guaranteed service is thus
\begin{align}
\label{SPrevenue}
R(\varepsilon_i)= R_P(\varepsilon_i) + R_{\text{NP},i}, i\in\{1,2\}.
\end{align}

In reality, offering free services to consumers often comes with a cost to the SPs, such as the cost of service and online platform creation and continued operations. Furthermore, we note that free online services profit from using consumer data and therefore encumber data processing related costs.  Let $C(v_i; \varepsilon_i)$ denote the cost of offering free services with privacy risk level $\varepsilon_i$. We model $C(v_i; \varepsilon_i)$ as sum of two non-negative costs: (i) $C_{\text{QoS}}(v_i)$ of providing services with QoS $v_i$; and (ii) $C_{P}(\varepsilon_i)$ as the processing (data analytics) cost of exploiting private data to the level of $\varepsilon_i$ such that
\begin{align}
\label{SPcost}
C(v_i; \varepsilon_i) = C_{\text{QoS}}(v_i) + C_P (\varepsilon_i) , i\in\{1,2\}.
\end{align}
Thus, via \eqref{SPrevenue} and \eqref{SPcost}, our model captures the fact that the benefit of using private data by each SP involves both cost and revenue.
\subsubsection{Consumer Model}
\begin{figure}[h]%
	\centering
	\subfloat[Classical hotelling model ]{{\includegraphics[width=6.6cm]{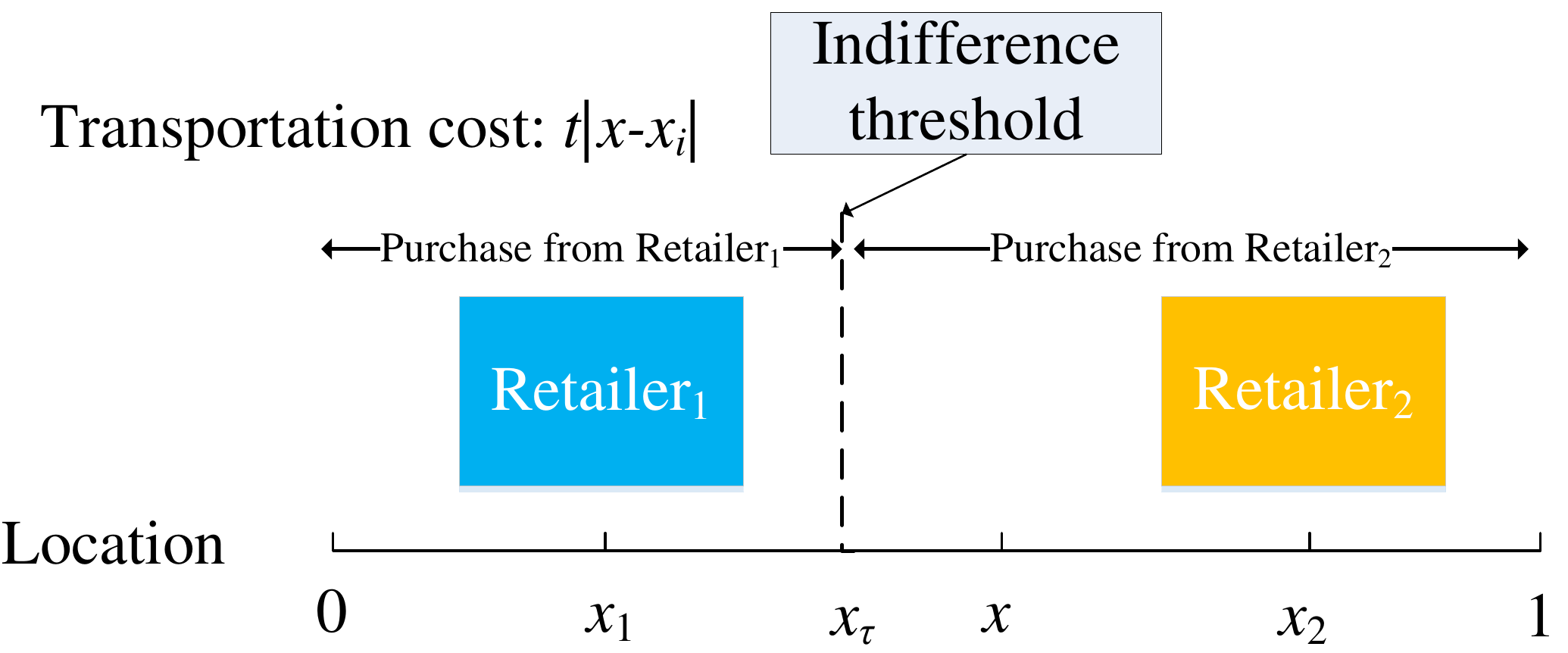}}\label{classicalhotelling}}%
	\quad
	\subfloat[Our modified hotelling model ]{{\includegraphics[width=6.6cm]{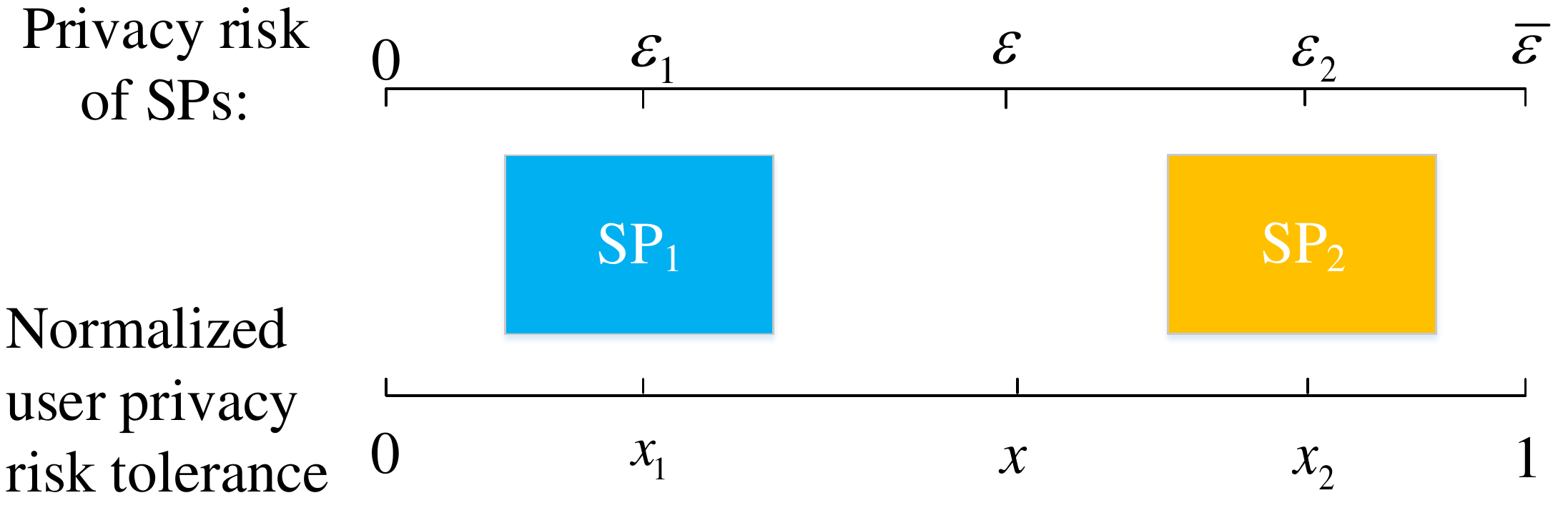}}\label{modifiedhotelling}}%
	\caption{User choice model for using different SPs}%
	\label{hotelling}%
\end{figure}

We build upon the classical Hotelling model to formulate both consumer utility and the resulting consumer-SP game. The Hotelling model maps retailers to two locations $(x_1, x_2)$ on a $[0,1]$ line such that the strategy of  each retailer is to determine the best location-price tuple that maximizes its profit. The location (see Figure~\ref{classicalhotelling}) is a proxy for a specific product differentiator. A consumer with its own product differentiator preference (traditionally assumed to be uniformly distributed over $[0,1]$) is mapped to a location $x\in[0,1]$ on the line as shown in Figure~\ref{classicalhotelling}. Such a spatial model allows computing the market segment by identifying both the optimal locations of the retailers and an indifferent threshold between the two optimal retailer locations at which both retailers are equally desirable. For such a uniform consumer preference model, the segmentation for each retailer is simply its distance to the indifference point (see Figure~\ref{classicalhotelling}). Consumers choose the retailer with the least product price and ``transportation cost" (modeled as a linear function of location) for a desired consumer valuation of the product. Note that transportation costs are metaphorical for any non-price-based differentiation of the two retailers.

For our problem, we obtain a Hotelling model by: (i) introducing a \textit{normalized privacy risk} and mapping it to spatial location; and (ii) by viewing the QoS as the net valuation of service by the consumer. Note that since we study a free services market, we use QoS as a measure of consumer satisfaction. We note that in the classical Hotelling model, the consumer pays a non-negative transportation cost for any retailer whose location is different from its own. However, our problem departs from this model in that higher and lower privacy risks offered by SPs relative to a consumer preferred privacy risk choice are not viewed similarly. 

We assume there exists infinitely many rational consumers that are interested in the services provided by the SPs. In keeping the standard game-theoretic definition, rational refers to consumers interested in maximizing some measure of utility via interactions with the SPs. We use a random variable $E\in[0,\bar{\varepsilon}]$ to denote the heterogeneous privacy choices of consumers; such a model assumes that the privacy preferences of consumers are independent and identically distributed, a reasonable assumption when the consumer set is very large. Let $E=\varepsilon$ denote the privacy risk preference of a consumer. If an $SP_i$ offers a privacy risk guarantee $\varepsilon_i$ higher than $\varepsilon$, then using its service will result in a privacy cost to the consumer due to perceived privacy risk violation. On the other hand, the consumer gains from choosing an $SP_i$ that offers an $\varepsilon_i<\varepsilon$ as a result of the extra privacy protection offered. Let $x=F_E(\varepsilon)\in[0,1]$ be a differentiable cumulative distribution function of $\varepsilon$. Thus, $x=F_E(\varepsilon)$ can be considered as a normalized privacy risk tolerance (i.e., restricted to $[0,1]$) which indicates the proportion of the consumers with a privacy risk tolerance of at most $\varepsilon$.  Since the privacy risk $\varepsilon_i$ can be over an arbitrary range $[0,\bar{\varepsilon}]$, the normalized spatial privacy risk is given by the CDF $F_E(\varepsilon_i)$.  Thus, for a consumer whose normalized privacy risk tolerance is located at $x\in [0,1]$, its actual privacy risk tolerance is $\varepsilon=F_E^{-1}(x)$. We can similarly map the privacy risks offered by the SPs to \textit{normalized locations} $x_1=F_E(\varepsilon_1)$ and $x_2=F_E(\varepsilon_2)$ on the $[0,1]$ line as shown in Figure~\ref{modifiedhotelling}.

Analogous to the Hotelling model, we let $u_i(x)$ denote the utility (in units of QoS) from $SP_i$ as perceived by a consumer with a normalized privacy preference (location) $x$. Our model for $u_i(x)$ contains two parts: (i) a positive QoS $v_i$ offered by $SP_i$; and (ii) the gain or loss in the perceived QoS as a result of a mismatch between consumer privacy preference and $SP_i$'s privacy risk offering. We introduce a gain factor $t$ that allows mapping the privacy mismatch $t(x-x_i)\varepsilon_i$ to a QoS quantity. This mismatch utility indicates that when the SP offers a service with privacy risk lower than the consumer's tolerance, the consumer receives a positive utility due to extra privacy protection. However, if the service offered has a higher privacy risk than the consumer's tolerance, the consumer will receive negative utility for privacy risk violation. In other words, given the same level of QoS, the better the privacy risk guarantee an SP offers, the more the consumer prefers the SP. We now write the utility or profit function for both consumers and SPs. 

\subsubsection{Consumer utility and SP profits} 
\label{subsection: twospvaluation}
For the consumer located at $x$, the overall perceived utility for choosing services provided by $SP_1$ and $SP_2$ are
\begin{align}
\label{uservaluation}
u_i(x)& =v_i+t(x-x_i)\varepsilon_i, i\in\{1,2\}.
\end{align} 
For each $SP_i, i\in\{1,2\}$, let $(v_{-i},\varepsilon_{-i})$ be its competitor's strategy. For the revenue and cost models in ~\eqref{SPrevenue} and \eqref{SPcost}, the profit of $SP_i$ is simply the difference
\begin{align}
\label{objpi}
\pi_i(v_i;\varepsilon_i; v_{-i}; \varepsilon_{-i})& =[R(\varepsilon_i)-C(v_i;\varepsilon_i)]n_i(v_i; \varepsilon_i; v_{-i}; \varepsilon_{-i}), i\in\{1,2\},
\end{align}
where $n_i(v_i; \varepsilon_i; v_{-i}; \varepsilon_{-i})$ denotes the fraction of consumers who choose $SP_i$.
\begin{assumption}
\label{positiveqos}
We assume that the services provided by both SPs have non-negative QoS. 
\end{assumption}
Since consumers are rational, they expect to have positive utility through the interactions with the SPs. It is reasonable to assume that SPs have no incentive to offer services with a negative QoS. In other words, we assume $v_1\ge0$ and $v_2\ge 0$.
\begin{assumption}
\label{completecoverage}
We assume the model parameters are chosen such that they ensure the market is completely covered by $SP_1$ and $SP_2$.
\end{assumption}
The above assumption implies that each consumer must choose one of the SPs. Such an assumption is implicitly built into the classical Hotelling model to ensure competition between SPs and our model continues to do so too. Later we provide a sufficient condition for sustaining the equilibrium market segmentation under these assumptions.
\subsection{Two-SP Non-cooperative Game Formulation}
We note that the SPs compete against each other through their distinct QoS and privacy risk offerings, which in turn affects consumer choices and helps determine the stable market segmentation. Thus, the interactions between $SP_1$ and $SP_2$ can be formulated as a non-cooperative game in which the strategy (action) of each SP is a (privacy risk, QoS) tuple and that of the consumer is choosing an SP. Furthermore, we assume that the SPs are rational and have perfect information, implying that they play to maximize their own profits and know the exact profit function for any given strategy.

The interactions between retailers and consumers in the Hotelling model can be viewed as a sequential game~\cite{hotelling1990stability}. For our model, such a sequential game involves three stages. In the first stage, the differentiator, i.e. the normalized privacy risk $\varepsilon_i$, is advertised by $SP_i$. Thus is followed by each SP determining its QoS for the advertised risk. Finally, the consumers choose the preferred SP based on the ($\varepsilon_i, v_i$) tuple that maximizes its utility. 

The game can be formally described as follows: (i) a set of players $\{1, 2, \mathcal{C}\}$, where $1$ and $2$ denote $SP_1$ and $SP_2$, respectively, and the set $\mathcal{C}$ contains infinitely many consumers; (ii) a collection of strategy tuples $(\varepsilon_i, v_i) \in \mathcal{E}_i\times\mathcal{V}_i$ for $SP_i$ and a collection of binary choices (strategies) for the consumer $b\in \mathcal{B}=\{1, 2\}$;
and (iii) a profit function $\pi_i$ for each $SP_i$ and a utility function $u_i$ for each consumer for choosing $SP_i$.

\subsection{The Subgame Perfect Nash Equilibrium for the Two-SP Game}
In a sequential game, each stage is referred to as a subgame~\cite{fudenberg1991game}. One often associates a strategy profile with a sequential game. A strategy profile is a vector whose $i^{\text{th}}$ entry is the strategy for all players at the $i^{\text{th}}$ stage of the sequential game. A non-cooperative sequential game has one well-studied solution: the Subgame Perfect Nash Equilibrium (SPNE). 

A strategy profile is an SPNE if its entries are the Nash equilibria of the subgame resulting at each stage of the sequential game. The SPNE of a sequential game captures an equilibrium solution such that no player can make more profit by unilaterally deviating from this strategy in every subgame. 

Since the above non-cooperative game is a game with finite number of stages and perfect information, it can be solved using backward induction. Backward induction is the process of reasoning backwards in time, starting from the last stage of the sequential game, to determine a sequence of optimal strategies. It proceeds by first determining the optimal strategies in the last stage. Using this information, one can then decide the optimal strategies for the second-to-last stage of the game. This process continues backwards until the optimal strategies for every stage has been determined. We apply backward induction to the three stage game described above as follows.

\textit{\textbf{Stage 3, Users' decisions:}} Each consumer located at $x\in[0,1]$ can choose the services provided by either $SP_1$ or $SP_2$ based on its valuation function in~\eqref{uservaluation}. The resulting optimal strategy for the consumer is to choose the SP whose index is given by
\begin{align}
\label{useropt}
arg\max_{i\in \{1, 2\}} v_i+t(x-x_i)\varepsilon_i.
\end{align}
Since the consumer's utility is a linear function of the normalized privacy risk $x$ and the market is completely covered by the SPs, there exists a threshold $x_\tau$ such that the consumer located at $x_\tau$ is indifferent to using services provided by $SP_1$ or $SP_2$. Thus, at the indifference threshold $x_\tau$, we have 
\begin{align}
& u_2(x_\tau)=u_1(x_\tau)\\\nonumber
\implies & v_2+t(x_\tau-x_2)\varepsilon_2=v_1+t(x_\tau-x_1)\varepsilon_1.
\end{align}
Simplifying further, the indifference threshold for choosing between the two SPs is given by
\begin{align}
\label{indifference}
x_\tau=\frac{v_1-v_2+t(F_E(\varepsilon_2)\varepsilon_2-F_E(\varepsilon_1)\varepsilon_1)}{t(\varepsilon_2-\varepsilon_1)},
\end{align} 
where $x_1$ and $x_2$ have been replaced by their corresponding normalized privacy risk values. Thus, given the SPs' tuples $(\varepsilon_i,v_i), i\in\{1,2\}$, the optimal strategy of a consumer located at $x$ is to use the service of $SP_1$ if $x\le x_\tau$ and $SP_2$ otherwise.

\textit{\textbf{Stage 2, SPs determine QoS:}}
In the second stage, for a given privacy risk guarantee $\varepsilon_i$, $SP_i$ chooses its QoS $v_i$ to maximize its profit $\pi_i$. Since a consumer's normalized privacy risk tolerance denotes the fraction of the population whose privacy risk tolerance is at most $\varepsilon$, $x_\tau$ determines the proportion of consumers who choose $SP_1$, i.e., $n_1$. As a result, the profit functions of $SP_1$ and $SP_2$ can be written as
\begin{align}
\label{pie}
\pi_1(v_1;\varepsilon_1; v_{2}; \varepsilon_{2})=& [R(\varepsilon_1)-C(v_1;\varepsilon_1)]\frac{v_1-v_2+t(F_E(\varepsilon_2)\varepsilon_2-F_E(\varepsilon_1)\varepsilon_1)}{t(\varepsilon_2-\varepsilon_1)},\\
\label{pii}
\pi_2(v_1;\varepsilon_1; v_{2}; \varepsilon_{2})=&  [R(\varepsilon_2)-C(v_2;\varepsilon_2)][1-\frac{v_1-v_2+t(F_E(\varepsilon_2)\varepsilon_2-F_E(\varepsilon_1)\varepsilon_1)}{t(\varepsilon_2-\varepsilon_1)}].
\end{align}

To find the SPNE in this stage, we use the best response method~\cite{osborne1994course}. The best response is a function which captures the behavior of each player while fixing the strategies of the other players. For any $v_{-i}\in \mathcal{V}_{-i}$, we define $BR_i(v_{-i})$ as the best strategy of $SP_i$ such that 
\begin{align}
BR_i(v_{-i})=arg\max\limits_{v_i}\pi_i(v_i;\varepsilon_i; v_{-i}; \varepsilon_{-i}), i\in\{1,2\}.
\end{align}In the Nash equilibrium, each player plays the best response with respect to other players' strategies. Thus, a Nash equilibrium in this stage is a profile $\vec{v}^*=(v^*_i, v^*_{-i})$ for which 
\begin{align}
v^*_i\in BR_i(v_{-i}), \forall i\in\{1,2\}.
\end{align}
To find the Nash equilibria, we first calculate the best response function of each SP, then find a strategy profile $\vec{v}^*$ for which $v^*_i\in BR_i(v_{-i}), \forall i\in\{1,2\}$. For a given set of privacy risk guarantees $\{\varepsilon_1, \varepsilon_2\}$, the optimal QoS $v^*_i$ of $SP_i, i \in\{1,2\}$ in the SPNE is then determined by the solution to the following set of simultaneous equations
\begin{align}
	\label{optD}
\begin{array}{l}
v^*_1=arg\max\limits_{v_1}\pi_1(v_1;\varepsilon_1; v_{2}; \varepsilon_{2}),\\
v^*_2=arg\max\limits_{v_2}\pi_2(v_1;\varepsilon_1; v_{2}; \varepsilon_{2}).
\end{array}
\end{align} 

\textit{\textbf{Stage 1, SPs determine privacy risk guarantee:}}
In the first stage, we compute equilibrium strategies $\varepsilon_1$ and $\varepsilon_2$ that the two SPs should advertise for optimal market share. Note that $x_\tau, v^*_1$, and $v^*_2$ have been computed in stages 1 and 2 for a fixed $\varepsilon_1$ and $\varepsilon_2$, and therefore, are functions of $\varepsilon_1$ and $\varepsilon_2$. The objective functions $\pi_1$ and $\pi_2$ are thus also functions of $\varepsilon_1$ and $\varepsilon_2$; this in turn implies they can be maximized to find the equilibrium strategy $\varepsilon^*_1$ and $\varepsilon^*_2$ using the best response method.


\section{Two-SP Market with Linear Cost and Revenue Functions}
\label{Section:2SPs}
Thus far, we have considered a general model for the consumer distribution of privacy preferences. To obtain better intuition and meaningful analytical solutions, we consider a linear cost and revenue model for each SP. We define the cost function of $SP_i$ to be 
\begin{align}
\label{clinear}
C(v_i; \varepsilon_i) =cv_i+c\lambda\varepsilon_i, i\in\{1,2\},
\end{align}
where $c$ and $\lambda$ are constant scale factors in units of cost/QoS and QoS/privacy risk, respectively. In addition, we model the revenue of each SP from offering a privacy guaranteed service by a linear function
\begin{align}
\label{vlinear}
R(\varepsilon_i) &= r\varepsilon_i+p_i, i\in\{1,2\},
\end{align}
where $r$ is the revenue per unit privacy risk for using consumers' private data. The parameters $p_1$ and $p_2$ model the fixed revenues of the SPs that are independent of consumers' private data.


\subsection{Consumers with Uniformly Distributed Privacy Risk Tolerance}
We assume consumers have uniformly distributed privacy risk tolerance between $0$ and $\bar{\varepsilon}$. The resulting normalized privacy risk of each SP is given by 
\begin{align}
\nonumber
x_i=F_E(\varepsilon_i)=\frac{\varepsilon_i}{\bar{\varepsilon}}, i\in\{1,2\}. 
\end{align}
Let 
\begin{align}
\alpha=\frac{r}{c}-\lambda
\end{align} 
and \begin{align}
\label{MPRC}
\tilde{C}=ct\bar{\varepsilon}.
\end{align}
Note that $\alpha$ is the ratio of net profit from using consumer data for a unit of privacy risk to the cost for providing a unit of QoS. Furthermore, $\tilde{C}$ is the cost of providing non-zero utility to the consumer with a maximal mismatch of privacy risk (relative to SP).

By using the backward induction method, the computed SPNE of the two-SP non-cooperative game is presented in the following theorem.
\begin{theorem}
	\label{Theorem:SPNE2SP}
There exists an SPNE for the two-SP non-cooperative game if the model parameters $\{c,\alpha,t,\bar{\varepsilon}, p_1, p_2\}$ satisfy
\begin{align}
\label{feasiblehotelling}
-1 &\le \frac{16(p_2-p_1)}{9ct\bar{\varepsilon}} \le 1,\\
\label{feasibleeps}
\frac{4\alpha-3t}{3t} &\le \frac{16(p_2-p_1)}{9ct\bar{\varepsilon}} \le \frac{4\alpha-t}{3t},\\
\label{mktcoverage}
(12c\alpha\bar{\varepsilon})^2-(15ct\bar{\varepsilon})^2&+ 288ct\bar{\varepsilon}(p_2+p_1)\ge [16(p_2-p_1)]^2.
\end{align}
The closed form solution of the SPNE is given by
\begin{align}
\label{optepsilinear}
\varepsilon^*_2& =\frac{12\bar{\varepsilon}c\alpha+15ct\bar{\varepsilon}-16(p_2-p_1)}{24tc},\\
\label{optvilinear}
v^*_2& =\frac{(2\alpha+t)c\alpha6\bar{\varepsilon}+(\alpha-t)9ct\bar{\varepsilon}+(t-2\alpha)8p_2+(\alpha+t)16p_1}{24ct},\\
\label{optepselinear}
\varepsilon^*_1&=\varepsilon^*_2-\frac{3\bar{\varepsilon}}{4},
\\
\label{optvelinear}
v^*_1& =v^*_2-\frac{3\bar{\varepsilon}}{4}\alpha+\frac{p_2-p_1}{3c}.
\end{align}
At the equilibrium, i.e., for the SPNE, $\frac{1}{2}-\frac{8(p_2-p_1)}{9ct\bar{\varepsilon}}$ of the population choose the service provided by $SP_1$ while the remaining  $\frac{1}{2}+\frac{8(p_2-p_1)}{9ct\bar{\varepsilon}}$ of the population choose $SP_2$. The total profit of $SP_1$ and $SP_2$ are given by
$$ \pi^*_1=\frac{4c}{27t\bar{\varepsilon}}(\frac{9t\bar{\varepsilon}}{8}-\frac{2(p_2-p_1)}{c})^2; \pi^*_2=\frac{4c}{27t\bar{\varepsilon}}(\frac{9t\bar{\varepsilon}}{8}+\frac{2(p_2-p_1)}{c})^2.$$
\end{theorem}

The proof of Theorem~\ref{Theorem:SPNE2SP} is provided in Appendix~\ref{Proof:SPNE2SP}.

In theorem~\ref{Theorem:SPNE2SP}, we observe $\varepsilon_2=\varepsilon_1+\frac{3}{4}\bar{\varepsilon}$, which implies for any SPNE strategy profile, the SP who with the higher QoS will offer a privacy risk guarantee $\frac{3}{4}\bar{\varepsilon}$ more than its competitor. Furthermore, a higher maximum net profit for using private data, denoted by $c\alpha\bar{\varepsilon}$, encourages SPs to offer higher privacy risk guarantees. However, a larger difference in revenues independent of private information between $SP_1$ and $SP_2$, denoted by $p_2-p_1$, encourages SPs to lower their privacy risk guarantees to attract more consumers with lower privacy risk tolerance. Also, a larger $\tilde{C}$ will result in a larger differentiation in privacy risk guarantee in the equilibrium strategies of $SP_1$ and $SP_2$. Finally, as $p_2-p_1$ increases, the market share of $SP_1$ decreases while $SP_2$'s market share increases.

%

By~\eqref{MPRC}, $x_\tau$, $\pi^*_1$, and $\pi^*_2$ can be simplified to
\begin{align}
&\quad x_\tau=\frac{1}{2}-\frac{8(p_2-p_1)}{9\tilde{C}}.\\
\pi^*_2=\frac{1}{3}(\frac{3}{4}\sqrt{\tilde{C}}+& \frac{4(p_2-p_1)}{3\sqrt{\tilde{C}}})^2 \quad\quad\pi^*_1=\frac{1}{3}(\frac{3}{4}\sqrt{\tilde{C}}-\frac{4(p_2-p_1)}{3\sqrt{\tilde{C}}})^2
\end{align}
Note that for a fixed $p_2-p_1$, both $\pi^*_1$ and $\pi^*_2$ are decreasing functions of $\tilde{C}$ when $\tilde{C}\in[0,\frac{16(p_2-p_1)}{9}]$ and increasing afterwards. On the other hand, by \eqref{feasiblehotelling}, we have $\frac{16(p_2-p_1)}{9}\le ct\bar{\varepsilon}$, which implies $\tilde{C}\ge\frac{16(p_2-p_1)}{9}$. Therefore, both $\pi^*_1$ and $\pi^*_2$ are increasing function of $\tilde{C}$ in the SPNE. This indicates both SPs will make more profits in the SPNE with a larger $\tilde{C}$.

\subsection{Consumers with Normally Distributed Privacy Risk Tolerance}
In this section, we model consumers' privacy tolerance as a random variable $E$ that follows a normal distribution $\mathcal{N}(\frac{\bar{\varepsilon}}{2}, \sigma^2)$ with a mean of $\frac{\bar{\varepsilon}}{2}$ and a standard deviation of $\sigma$. Since $E\in[0,\bar{\varepsilon}]$, we restricted the normal distribution to lie within the interval $[0,\bar{\varepsilon}]$. Then $E$ conditional on $E\in[0,\bar{\varepsilon}]$ follows a truncated normal distribution with cumulative distribution function
\begin{align}
\label{truncatednormal}
F_E(\varepsilon)=
\left\{
\begin{array}{ll}
\frac{\Phi(\frac{\varepsilon-\frac{\bar{\varepsilon}}{2}}{\sigma})-\Phi(-\frac{\bar{\varepsilon}}{2\sigma})}{\Phi(\frac{\bar{\varepsilon}}{2\sigma})-\Phi(-\frac{\bar{\varepsilon}}{2\sigma})} & \varepsilon\in[0,\bar{\varepsilon}] \\
0 & \varepsilon\in[-\infty,0]\\
1 & \varepsilon\in[\bar{\varepsilon},+\infty]\\
\end{array}
\right., 
\end{align}
where $\Phi(y)$ denotes the CDF of the standard normal distribution.

In contrast to the uniform distribution case, the CDF in \eqref{truncatednormal} is not amenable to a closed form solution. Thus, we characterize the equilibrium numerically. To find the SPNE,  we first compute the SPNE QoS in the second stage as functions of privacy risk guarantees by solving~\eqref{optD}. Then, we use an iterated best response method to find the optimal privacy risk guarantee of an SP by fixing other SPs strategies in each iteration. When the process converges, we have found an SPNE in which no SP is better off by unilaterally deviating from the equilibrium.

\subsection{Illustration of Results}
In this section, we illustrate our model and results. First, we assume consumers have uniformly distributed privacy risk tolerance. We plot each SP's SPNE strategies, market share, and total profit with respect to consumers' maximum privacy risk tolerance for different values of the QoS/privacy risk scale factor $t$. Later, we study the model in which consumers' privacy risk tolerance follows a normal distribution $\mathcal{N}(\frac{\bar{\varepsilon}}{2},1)$  truncated between $0$ and $\bar{\varepsilon}$. The model parameters are given as follows: 
\begin {table}[h]
\begin{center}
	\begin{tabular}{|c||c|c|c|c|c|}
		\hline
		Parameter                        & $c$  & $\lambda$ & 	$r$ & $p_1$ & $p_2$\\\hline
		Value                        &  0.5  & 0.75 & 0.7  &  0.4  & 0.8  \\\hline	                        
	\end{tabular}
	\caption{Numerical Example Model Parameters}
	\label{simulationparameters}	
\end{center}
\end{table}
\subsubsection{Consumers with Uniformly Distributed Privacy Risk Tolerance}
\begin{figure}[h]%
    \centering
    \subfloat[Privacy Risk of SPs vs. consumers' maximum privacy risk tolerance for different values of the QoS/privacy risk scale factor $t$ at SPNE]{{\includegraphics[width=60.5mm]{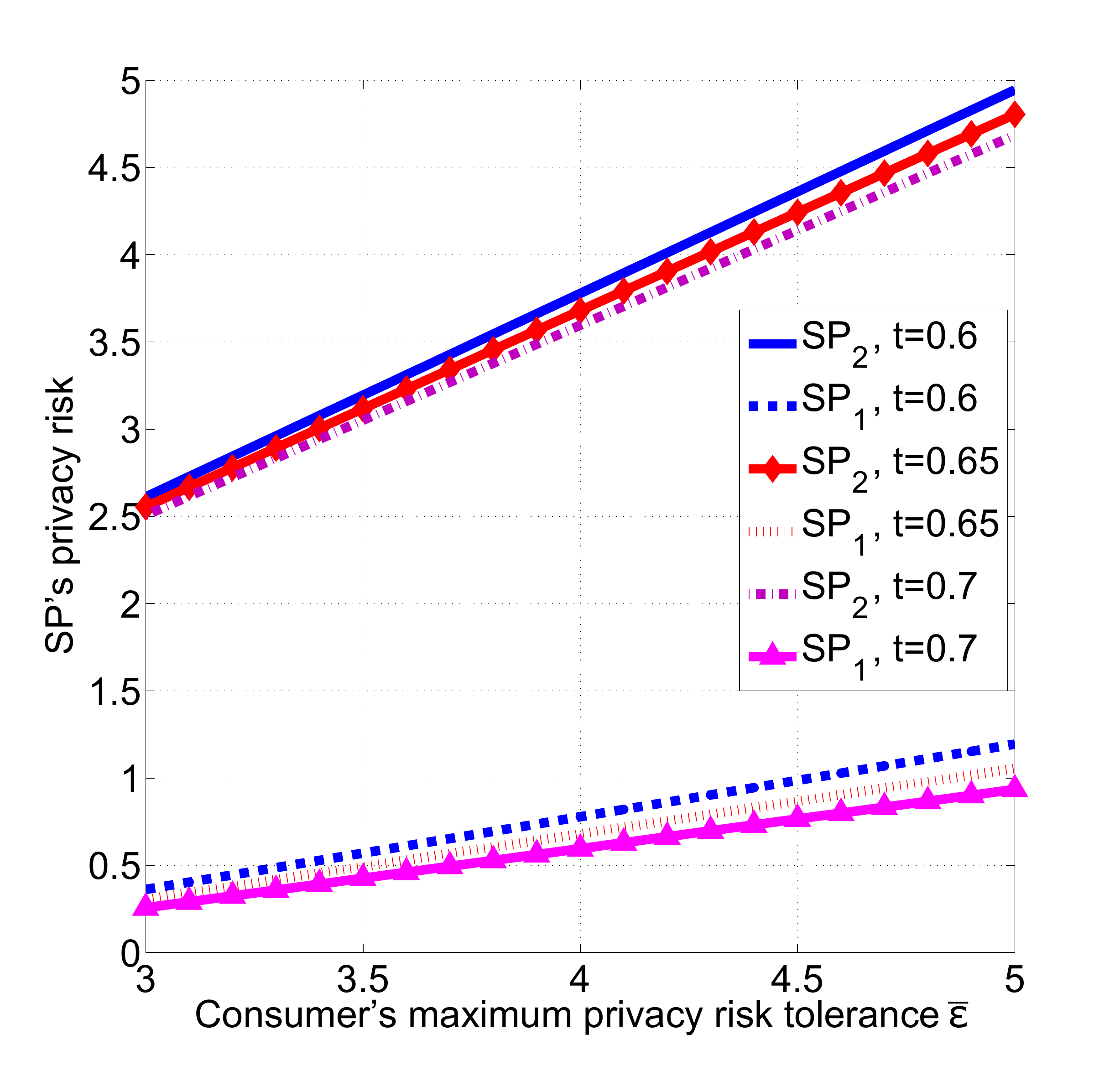} }\label{UVaryingtPrivacyRisk}}%
    \qquad
    \subfloat[QoS of SPs vs. consumers' maximum privacy risk tolerance for different values of the QoS/privacy risk scale factor $t$ at SPNE]{{\includegraphics[width=61.5mm]{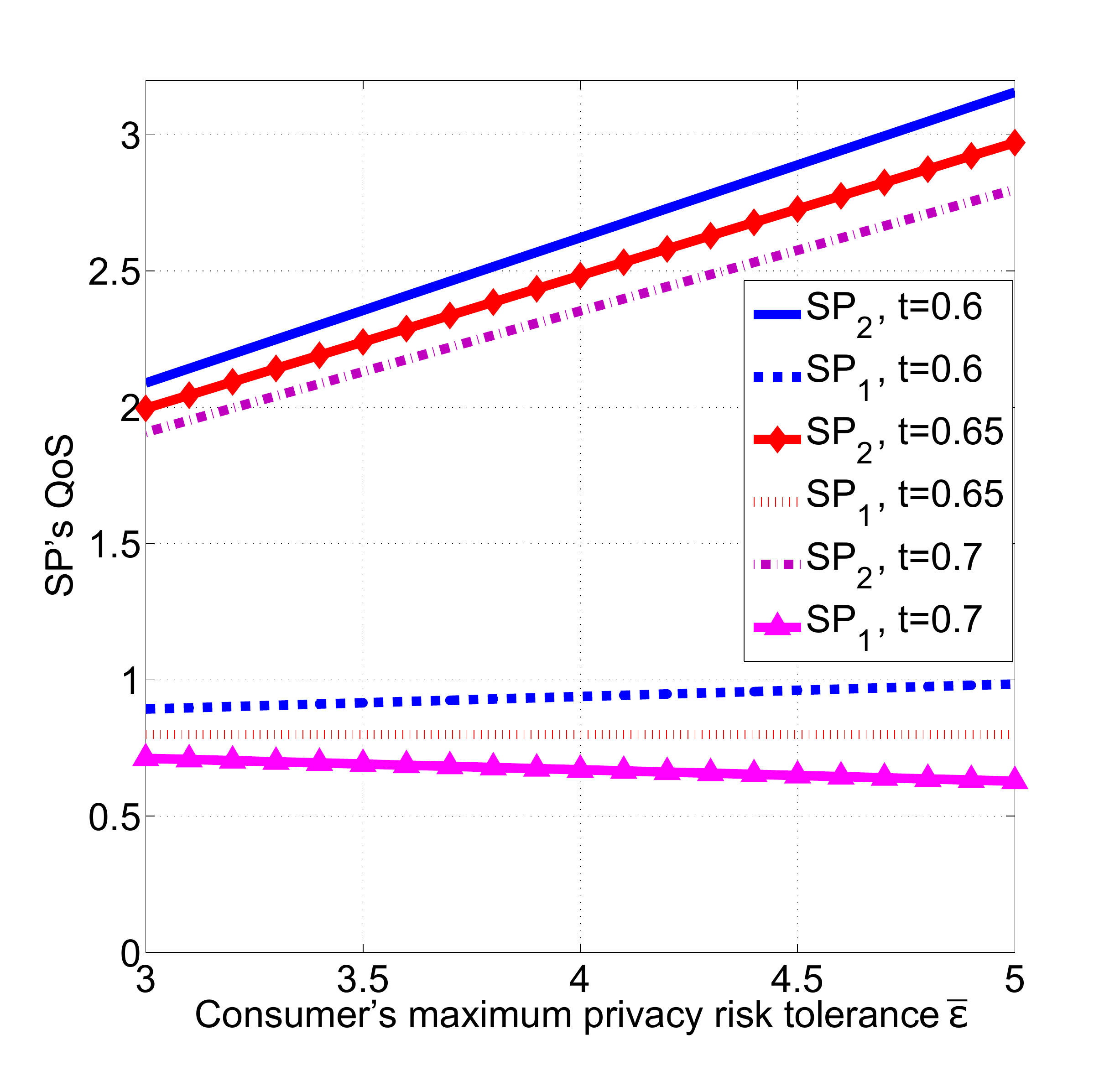} }\label{UVaryingtQoS}}%
    \caption{Equilibrium strategies of SPs vs. consumers' maximum privacy risk tolerance for different values of the QoS/privacy risk scale factor $t$ under a uniform distribution of consumer privacy risk}%
    \label{fig:equilibriumstrategiesU}%
\end{figure}

In this section, we vary $\bar{\varepsilon}$ from $3$ to $5$ to study properties of SPNE. Note that by theorem~\ref{Theorem:SPNE2SP}, $t$ must belong to $[0.58, 0.85]$ for a stable and sustainable SPNE. In Figure~\ref{fig:equilibriumstrategiesU}, we plot the equilibrium strategies of different SPs. Observe that both privacy risks and QoSs are linear functions of $\bar{\varepsilon}$, as expected for the linear model assumed. Furthermore, it can be seen that as $t$, the valuation of privacy by consumer, decreases, each SP will increase its privacy risk to generate more profit from using private data. Correspondingly, the SPs will have to provide higher QoS to attract consumers. 

\begin{figure}[h]
 		\centering
 		\includegraphics[width=10cm]{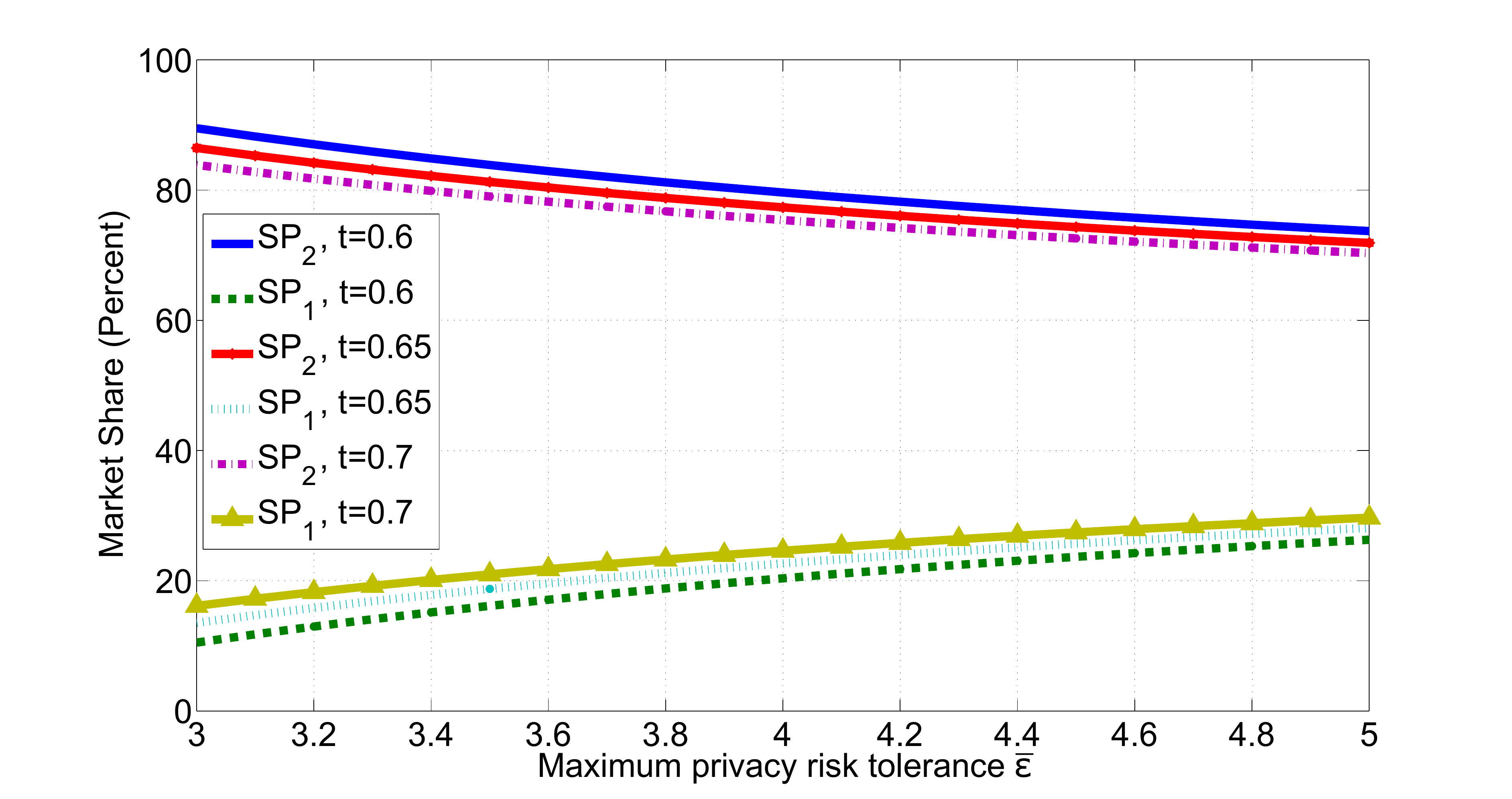}
 		\caption{Market shares of SPs at SPNE vs. consumers' maximum privacy risk tolerance for different values of the QoS/privacy risk scale factor $t$ under a uniform distribution of consumer privacy risk}
 		\label{fig:marketshareuniform}
 \end{figure}

The market shares of different SPs in the SPNE are presented in Figure~\ref{fig:marketshareuniform}. We observe that the equilibrium market share of $SP_2$ decreases as $t$ increases. The intuition behind this is that if $t$ increases, the consumer's valuation of privacy mismatch also increases. Thus, it is more difficult for $SP_2$ to attract consumers with privacy tolerance lower than $\varepsilon_2$. As a result, its market share decreases.  Notice that in Figure~\ref{fig:marketshareuniform}, as $\bar{\varepsilon}$ decreases, the equilibrium market share of $SP_2$ also increases. This is because consumers experience a smaller negative utility from the mismatch between their preferred and the offered privacy risk when the net range is smaller (recall that the utility from mismatch is given by $t(x-x_i)\varepsilon_i, \varepsilon_i\in[0,\bar{\varepsilon}]$). Thus, $SP_2$ has the incentive to offer high QoS with high privacy risk. As a result, more consumers will choose the SP with a higher privacy risk to enjoy a higher QoS. 

\begin{figure}[h]%
    \centering
    \subfloat[Total profit of $SP_2$ vs. consumers' maximum privacy risk tolerance for different values of the QoS/privacy risk scale factor $t$ at SPNE ]{{\includegraphics[width=60.5mm]{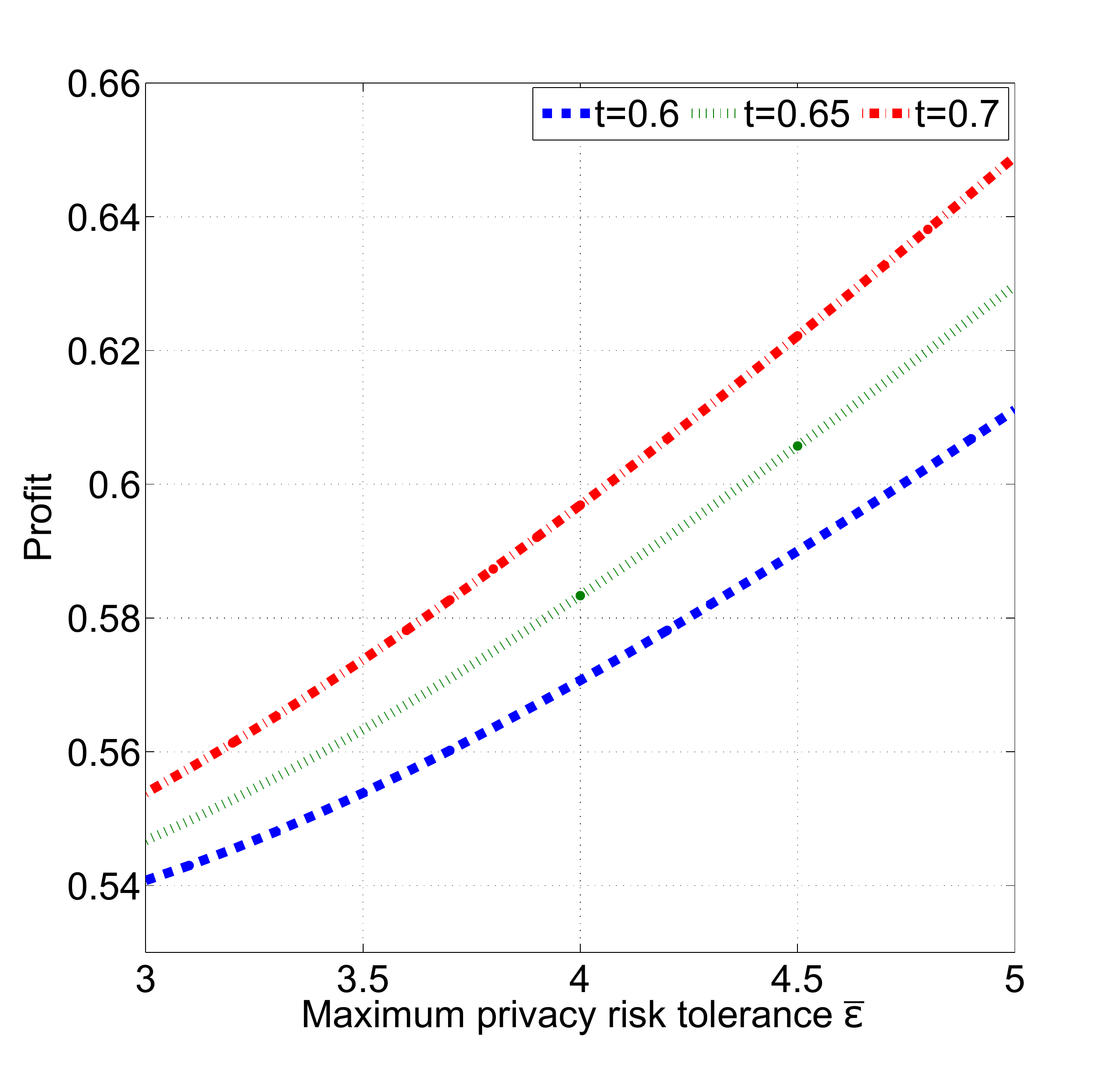} }\label{UVaryingtProfitSPA}}%
    \qquad
    \subfloat[Total profit of $SP_1$ vs. consumers' maximum privacy risk tolerance for different values of the QoS/privacy risk scale factor $t$ at SPNE]{{\includegraphics[width=61.5mm]{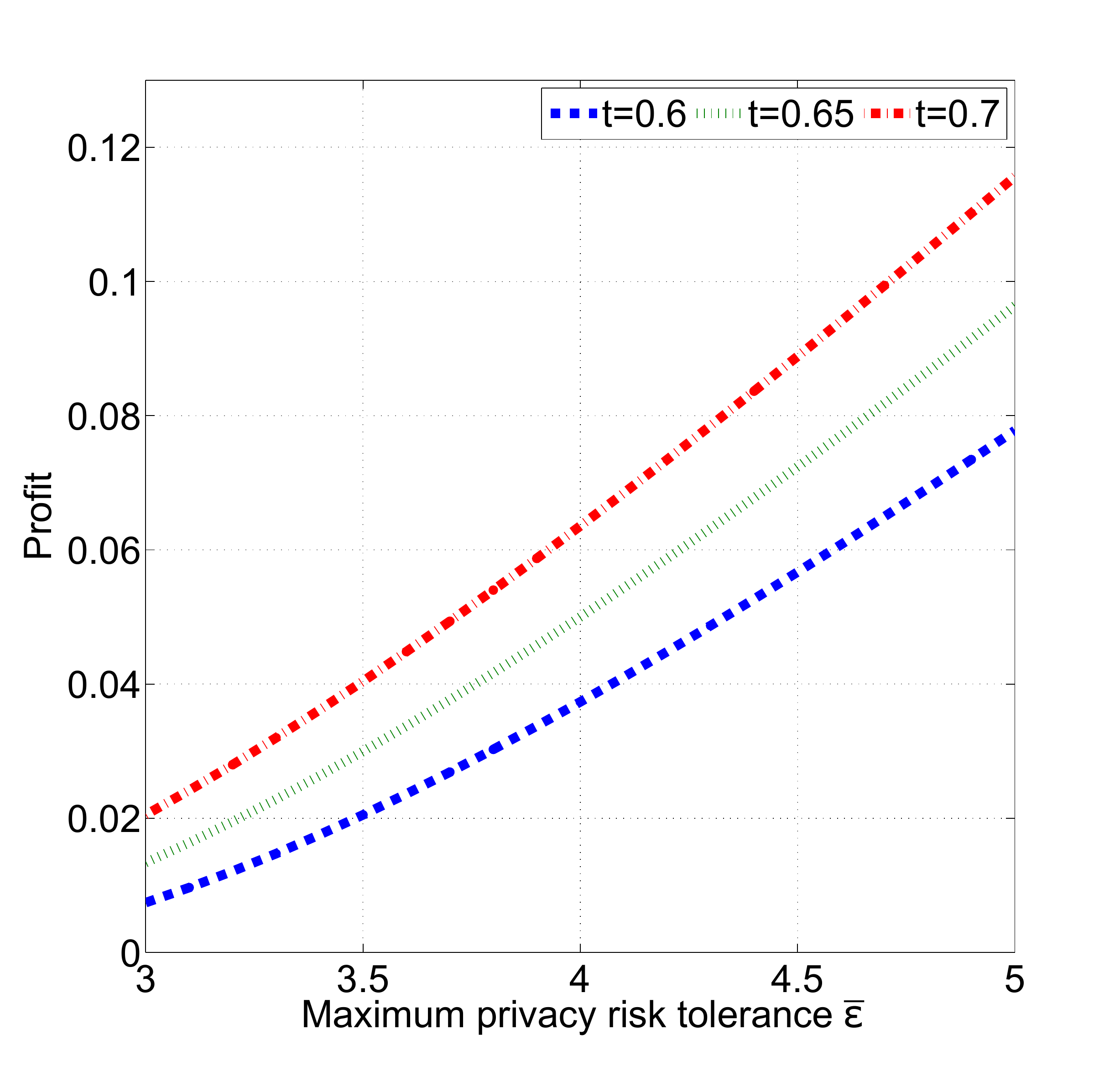} }\label{UVaryingtProfitSPB}}%
    \caption{Total equilibrium profit of SPs vs. consumers' maximum privacy risk tolerance for different values of the QoS/privacy risk scale factor $t$ under a uniform distribution of consumer privacy risk}%
    \label{fig:totalprofituniform}%
\end{figure}

In Figure~\ref{fig:totalprofituniform}, we plot the total profit at the SPNE for each SP as a function of the maximum consumer privacy risk tolerance $\bar{\varepsilon}$ for different values of $t$. As shown in the figure, the total profit of both SPs at SPNE increases as $\bar{\varepsilon}$ increases. This is due to the fact that a larger $\bar{\varepsilon}$ indicates a larger range of consumer preferences, and then, more possibilities for the SPs to exploit private information. Thus, both SPs can benefit from the using the private data of consumers that have a higher privacy risk tolerance. As $t$ decreases, the total profit of both SPs decrease. This is due to the fact that as $t$ decreases, the SPs intend to increase their privacy risks to generate more profit, As a result, the QoS of each SP increases to attract more consumers (see Figure~\ref{fig:equilibriumstrategiesU}), that in turn has the consequence of increasing the cost of providing services. Thus, the SPs have lower profits. 

\subsubsection{Consumers with Normally Distributed Privacy Risk Tolerance}
\begin{figure}[h]%
    \centering
    \subfloat[Privacy Risk of SPs consumers' maximum privacy risk tolerance for different values of $t$ at SPNE ]{{\includegraphics[width=60.5mm]{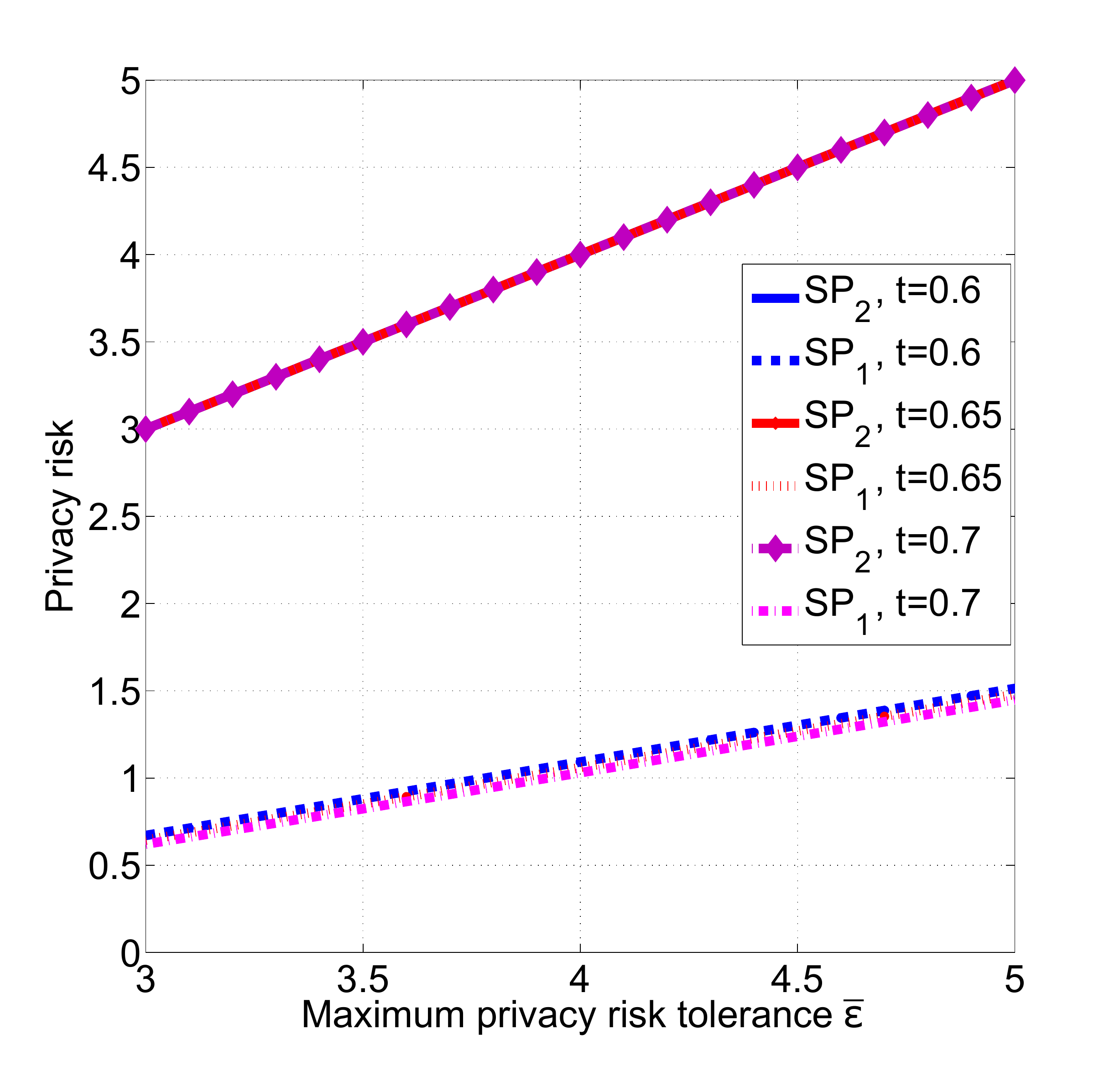} }\label{GVaryingtPrivacyRisk}}%
    \qquad
    \subfloat[QoS of SPs vs. consumers' maximum privacy risk tolerance for different values of $t$ at SPNE]{{\includegraphics[width=61.5mm]{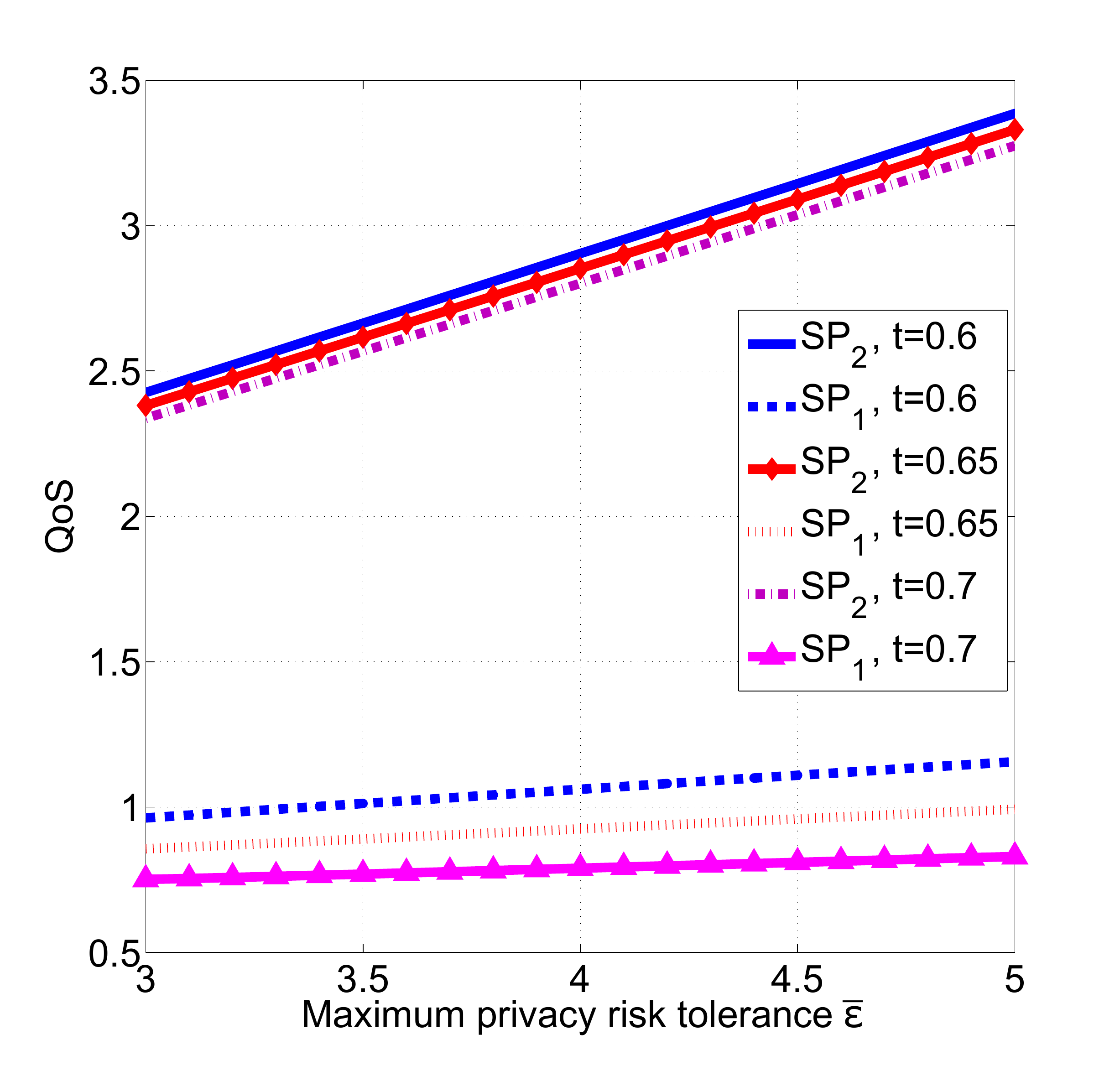} }\label{GVaryingtQoS}}%
    \caption{Equilibrium strategies of SPs vs. consumers' maximum privacy risk tolerance for different values of $t$ under truncated normal privacy risk tolerance distribution}%
    \label{fig:equilibriumstrategiesG}%
\end{figure}

We now consider the case in which consumers' privacy risk tolerance follows a truncated normal distribution with a mean of $\frac{\bar{\varepsilon}}{2}$ and a standard deviation of $1$. The equilibrium strategies of different SPs are shown in Figure~\ref{fig:equilibriumstrategiesG}. As with the uniform distribution scenario, here too we observe that the privacy risk and the QoS offered by each SP are linear functions of $\bar{\varepsilon}$. We also notice that in the SPNE, $SP_2$ will always provide service with maximum privacy risk (Figure~\ref{GVaryingtPrivacyRisk}). This is because the truncated normal distribution forces $SP_2$ to forfeit privacy differentiation while maximizing its profit. This allows $SP_1$ to gain an advantage (relative to the uniform distribution). Furthermore, we observe that the value of $t$ has slightly less influence on strategies of SPs in the SPNE compared to uniformly distributed consumer privacy tolerance. 

\begin{figure}[h]
	\centering
	\includegraphics[width=10cm]{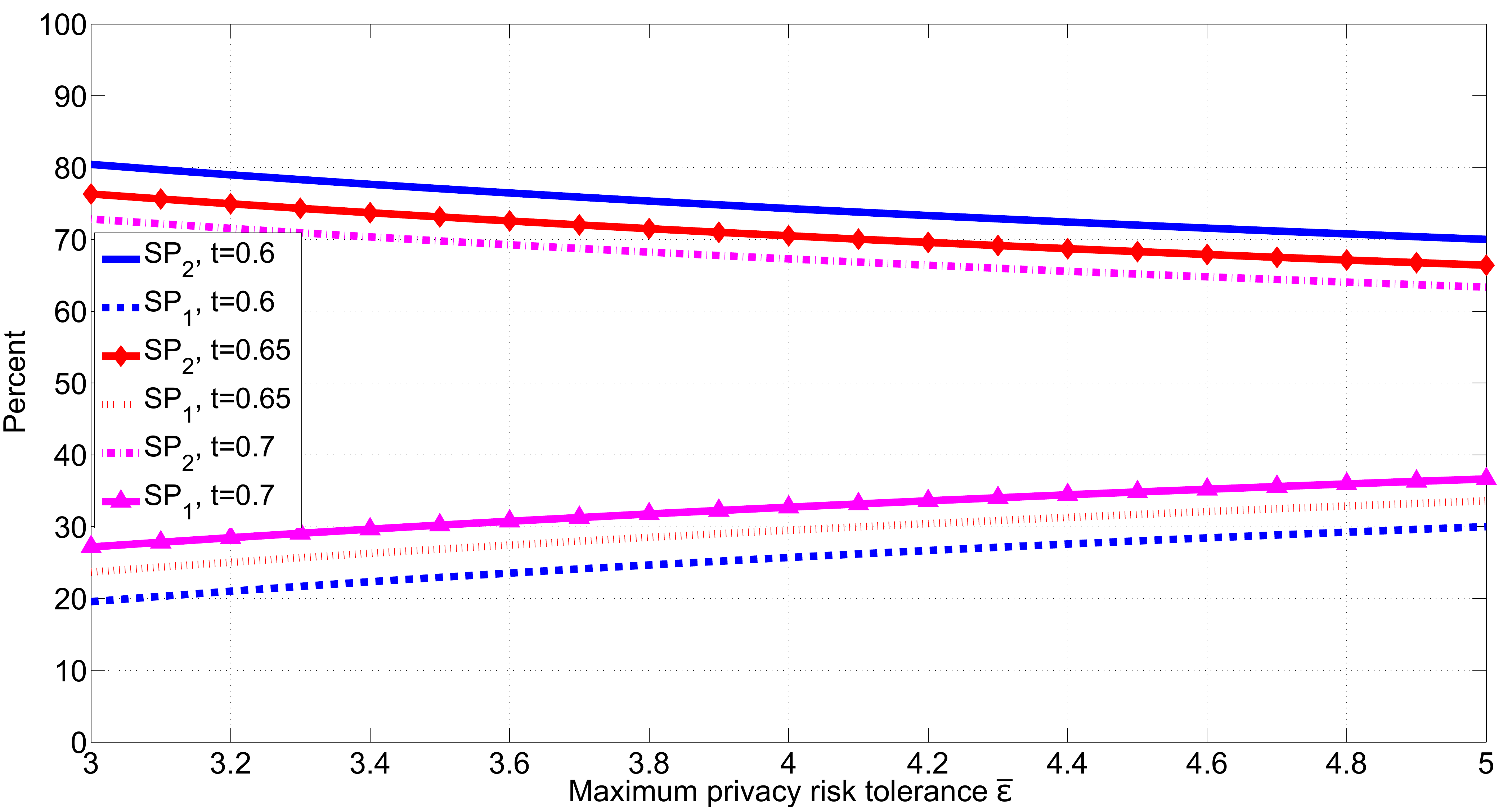}
	\caption{Market shares of SPs at SPNE vs. consumers' maximum privacy risk tolerance for different values of $t$ under truncated normal privacy tolerance distribution}
	\label{fig:marketshareG}
\end{figure}

Figure~\ref{fig:marketshareG} shows market shares of different SPs at SPNE vs. consumers' maximum privacy risk tolerance for different values of $t$ under truncated normal privacy tolerance distribution. As $t$ decreases, the market share of $SP_2$ at SPNE increases, and vice versa. Also, when $\bar{\varepsilon}$ decreases, the equilibrium market share of $SP_1$ also decreases. Furthermore, it can be seen that for the same $\bar{\varepsilon}$, the market share of $SP_2$ ($SP_1$) is smaller (larger) when consumers' privacy tolerance follows the truncated normal distribution compared to uniform distribution. Furthermore, our numerical analysis shows that at SPNE, $SP_2$ is forced to provide service with maximum privacy risk. We argue that this is due to the shape of the distribution that limits the number of consumers at the two extremes thus compelling the two SPs compete for the large bulk of consumers distributed around $\bar{\varepsilon}/2$. Given the ability of $SP_2$ to make more profit on untargeted services relative to $SP_1$, the SPNE solution leads to $SP_1$ increasing its market share to be profitable and $SP_2$ achieving profitability with a smaller market share.  

\begin{figure}[h]%
	\centering
	\subfloat[Total profit of $SP_2$ vs.consumers' maximum privacy risk tolerance for different values of $t$ at SPNE ]{{\includegraphics[width=60.5mm]{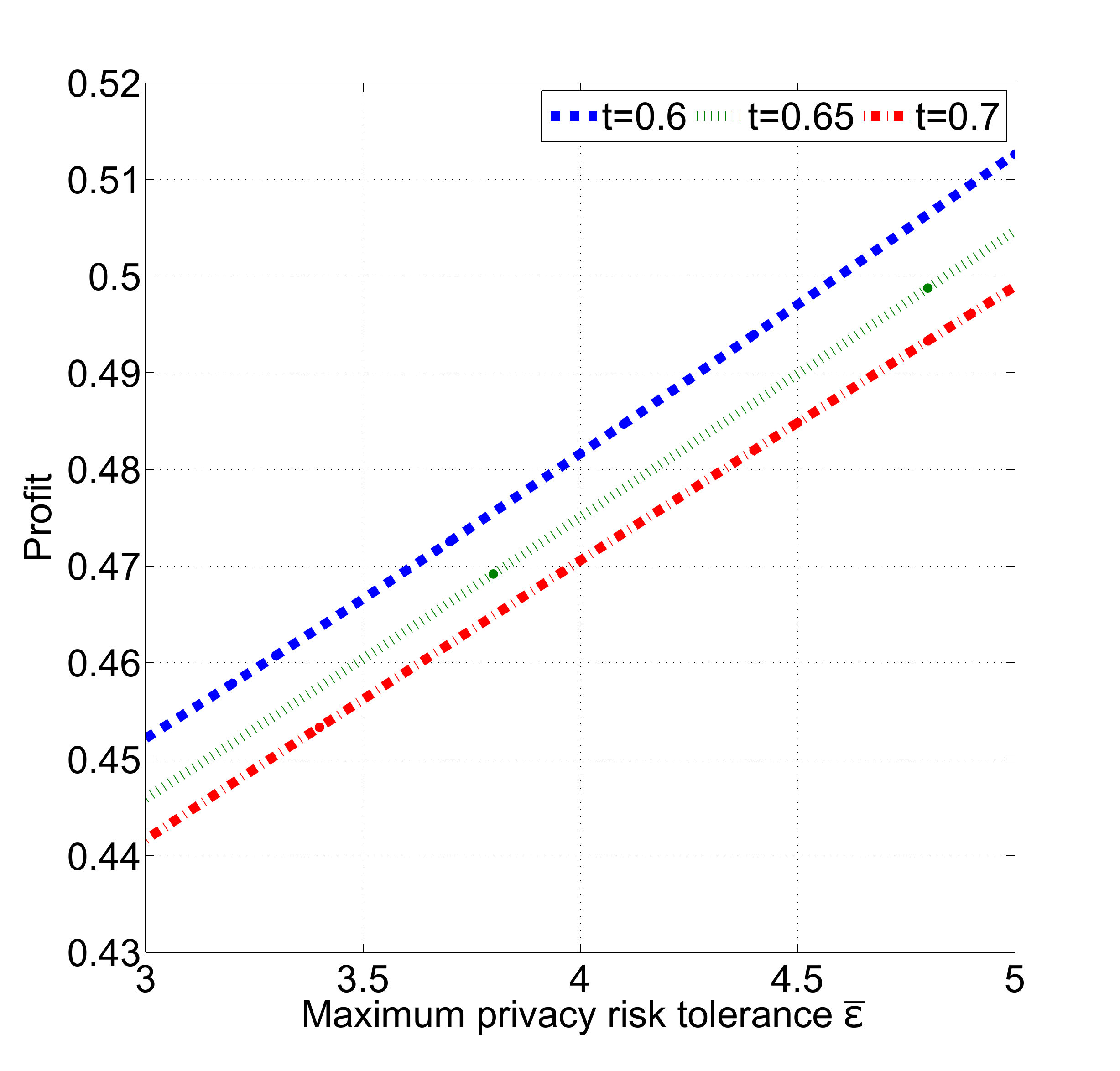} }\label{GVaryingtProfitSPA}}%
	\qquad
	\subfloat[Total profit of $SP_1$ vs. consumers' maximum privacy risk tolerance for different values of $t$ at SPNE]{{\includegraphics[width=61.5mm]{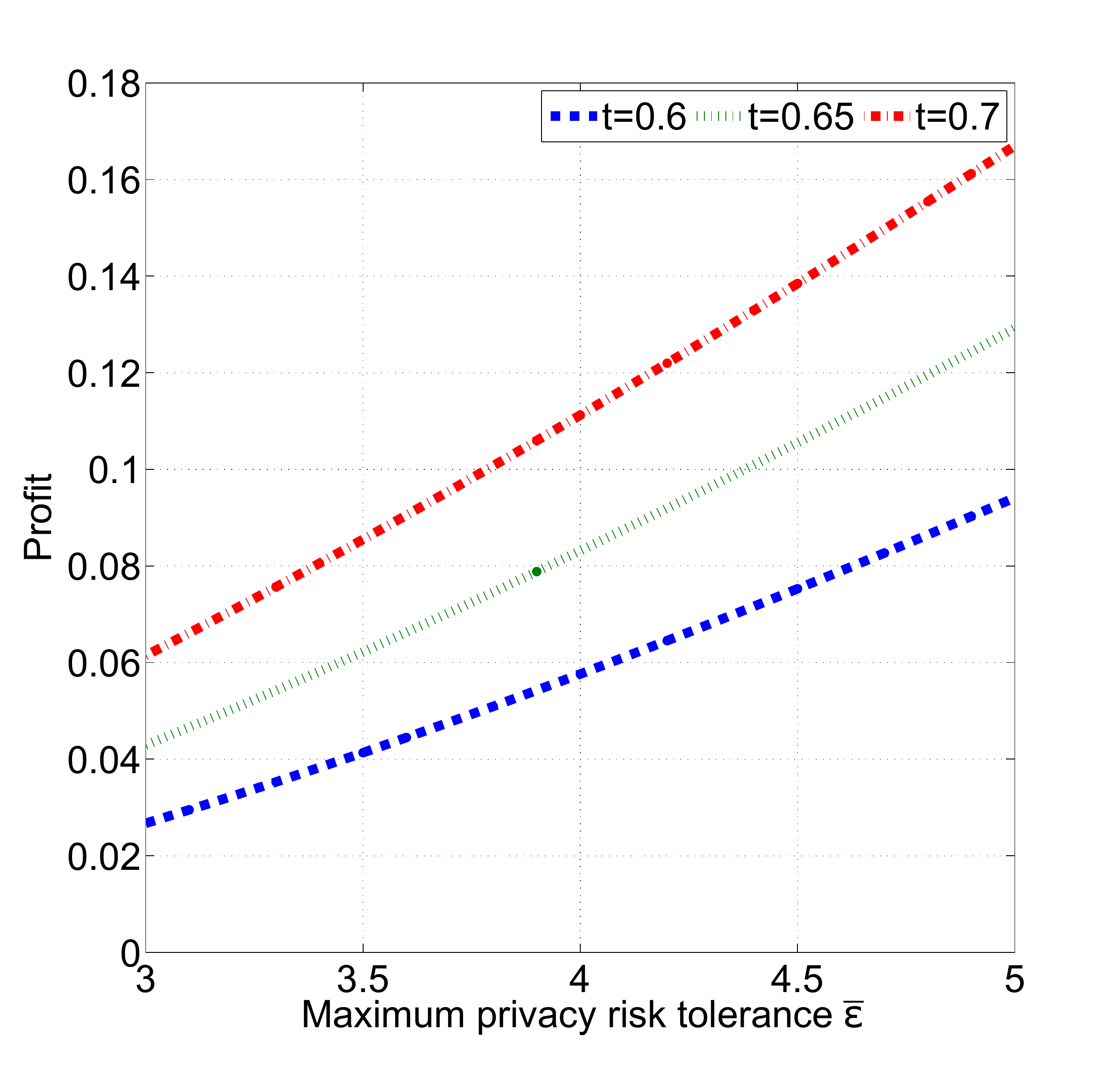} }\label{GVaryingtProfitSPB}}%
	\caption{Total equilibrium profit of SPs vs. consumers' maximum privacy risk tolerance for different values of $t$ under truncated normal privacy risk tolerance distribution}%
	\label{totalprofitG}%
\end{figure}

The relationship between total profit of different SPs at SPNE vs. consumers' maximum privacy risk tolerance for different values of $t$ is shown in Figure~\ref{totalprofitG}. Similar to Figure~\ref{fig:totalprofituniform}, both SPs' total profit increases as $\bar{\varepsilon}$ increases. However, in contrast to Figure~\ref{fig:totalprofituniform},  as $t$ decreases, the total profit of $SP_2$ increases. This is because $SP_2$ always offers $\bar{\varepsilon}$ in the SPNE. Notice that $SP_2$'s equilibrium QoS is also a linear function of $\varepsilon$ (see Figure~\ref{GVaryingtQoS}). On the other hand, $SP_2$'s market share increases as $t$ decreases (see Figure~\ref{fig:marketshareG}). By~\eqref{objpi}, \eqref{clinear}, and \eqref{vlinear}; the total profit of $SP_2$  increases as $t$ decreases.

\section{Market with Multiple Service Providers}
\label{Section:Multiple SPs}

\begin{figure}[h]
	\centering
	\includegraphics[width=14cm]{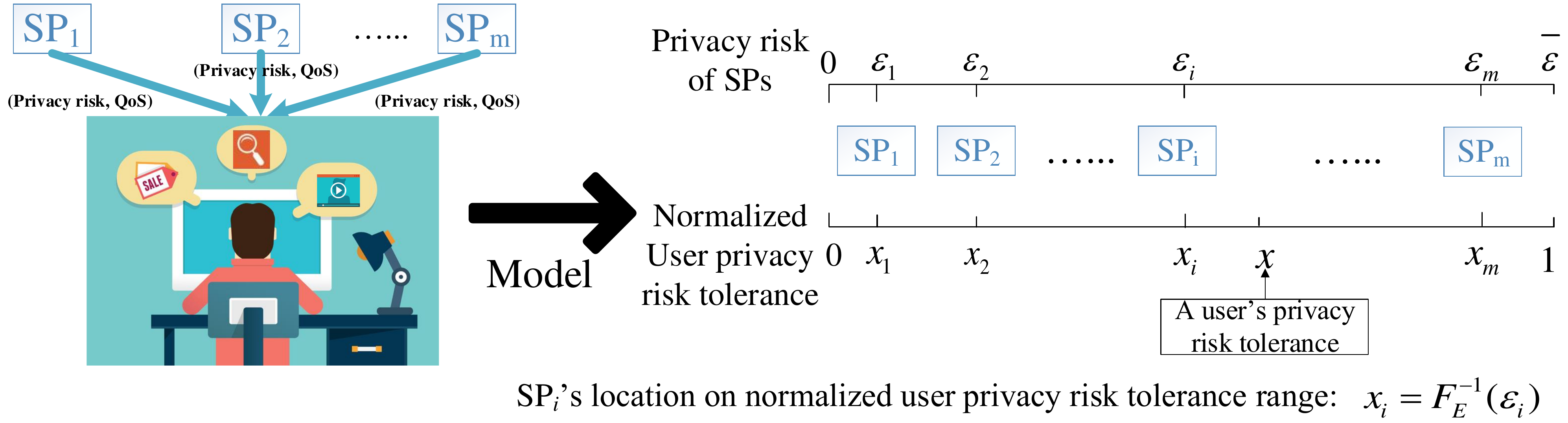}
	\caption{Market model for multiple SPs offering services with privacy guarantee}
	\label{fig:multispmodel}
\end{figure}

In the previous sections, we studied the market with two SPs. In this section, we examine a generalized model with multiple SPs (Figure~\ref{fig:multispmodel}). We allow for a finitely arbitrary number of SPs, each of which offers the same type of free service but with different QoS and privacy risk guarantee to consumers. In particular, we assume there are $m$ SPs in the market. Our models for cost, revenue and utility for each SP as well as the consumers are the same as for the two-SP model described in Section~\ref{subsection: twospvaluation}. Furthermore, we assume a consumer's privacy risk tolerance is uniformly distributed between $[0,\varepsilon]$. Analogous to the two SP model, the interactions between the $m$ SPs and consumers can also be viewed as a non-cooperative sequential game. The m-SP game proceeds in three stages. 

In the first stage, each of the $m$ SPs chooses its own privacy risk guarantee resulting in a vector $\vec{\varepsilon}=(\varepsilon_1,\varepsilon_2,...,\varepsilon_m)$ (on the interval $[0,\bar{\varepsilon}]$). Without loss of generality, we assume $\varepsilon_1\le\varepsilon_2\le...\le\varepsilon_m$. At the second stage, given the privacy risk $\vec{\varepsilon}$ determined in the first stage, the SPs simultaneously determine their QoS values to obtain a vector $\vec{v}=\{v_1,v_2,...,v_m\}$. At the last stage, each consumer chooses the SP that yields the maximal perceived utility for the consumer.   

To find the SPNE, we apply backward induction to the three stage game described above as follows. In the last stage of the game, for fixed QoS and privacy risk guarantee strategies of the SPs, consumers' choices of SPs are determined by their privacy risk tolerances. In the two-SP case, the consumer located at $x_\tau$ divides the set of consumers into two convex subsets where the consumers in the left subset will choose $SP_1$ and vice versa. However, for the multiple SP case, the market share of $SP_i$ ($i\in\{1,...,m\}$) is not necessarily a convex set between the  indifference threshold in which consumers are indifferent to choosing $SP_{i-1}$ or $SP_{i}$ and the threshold in which consumers are indifferent to choosing $SP_{i}$ or $SP_{i+1}$. This is due to the fact that in general the problem requires each $SP_i$ to compete with all other SPs, even if their privacy risk offerings are very different (e.g., SPs with a large difference in locations in Figure~\ref{fig:multispmodel}). We note that this will not happen in the equilibrium since an SP with zero market share would be better off by either improving its QoS to attract some consumers or just exit the market. Therefore, in the equilibrium, each SP only competes directly with its two closest neighbors. For given QoS profile $\vec{v}=\{v_1,v_2,...,v_m\}$ and privacy risk profile $\vec{\varepsilon}=(\varepsilon_1,...,\varepsilon_m)$, the market share of each SP are
\begin{align}
\nonumber
& n_1=\frac{v_1-v_2+t(F_E(\varepsilon_2)\varepsilon_2-F_E(\varepsilon_1)\varepsilon_1)}{t(\varepsilon_2-\varepsilon_1)},\\\nonumber
& n_i=\frac{v_i-v_{i+1}+t(F_E(\varepsilon_{i+1})\varepsilon_{i+1}-F_E(\varepsilon_{i})\varepsilon_i)}{t(\varepsilon_{i+1}-\varepsilon_i)}-\frac{v_{i-1}-v_i+t(F_E(\varepsilon_{i})\varepsilon_i-F_E(\varepsilon_{i-1})\varepsilon_{i-1})}{t(\varepsilon_i-\varepsilon_{i-1})}, \\\nonumber & \quad\quad\qquad i\in \{2,...,m-1\},\\\nonumber
& n_m=1-\frac{v_{m-1}-v_m+t(F_E(\varepsilon_{m})\varepsilon_N-F_E(\varepsilon_{m-1})\varepsilon_{m-1})}{t(\varepsilon_m-\varepsilon_{m-1})}.
\end{align}
Furthermore, we define the objective functions of $SP_i$ to be	
\begin{align}
\nonumber
 \pi_i(\vec{\varepsilon};\vec{v})=[R(\varepsilon_i)-C(v_i;\varepsilon_i)]n_i(\vec{\varepsilon};\vec{v}), i\in\{1,...,m\}.
\end{align}
For a given privacy risk guarantees profile $\vec{\varepsilon}$, the optimal QoS of $SP_i$ ($i\in\{1,...,m\}$) is determined by
\begin{align}
	\label{optmv}
	arg\max\limits_{v_i}\pi_i(\vec{\varepsilon};\vec{v}), i\in\{1,...,m\}
\end{align}
while fixing all other players' strategies.

We note that the cost function $C(v_i; \varepsilon_i)$ and the market segmentation computed in the first stage are both linear functions of $v_i$. Thus, for a fixed privacy risk guarantee profile $\vec{\varepsilon}$, the objective functions of $SP_i$ in this stage is a concave function with respect to its own strategy $v_i$. Furthermore, the feasible set of each SP's strategy is a convex set. Thus, the non-cooperative game among the SPs in this stage can is a m-player concave game. By~\cite{rosen1965existence}, there exists a Nash equilibrium. We define $\delta_i\triangleq\frac{1}{t(\varepsilon_{i+1}-\varepsilon_{i})}$,  $y_1\triangleq r(\varepsilon_1)+p_1-c\lambda\varepsilon_1-ctx_2\varepsilon_2+ctx_1\varepsilon_1$, $y_N\triangleq r(\varepsilon_N)+p_N-c\lambda\varepsilon_N-ct(1-x_N)\varepsilon_N+ct(1-x_{N-1})\varepsilon_{N-1}$ and $y_i\triangleq \frac{r(\varepsilon_i)+p_i-c\lambda\varepsilon_i+ctx_i\varepsilon_i-ctx_{i+1}\varepsilon_{i+1}}{t(\varepsilon_{i+1}-\varepsilon_i)}+\frac{r(\varepsilon_i)+p_i-c\lambda\varepsilon_i-ctx_{i-1}\varepsilon_{i-1}+ctx_i\varepsilon_i}{t(\varepsilon_{i}-\varepsilon_{i-1})} \quad \forall i\in\{2,...,m\}$. Applying the first order condition to SPs' profit functions (solving simultaneous linear equations obtained from $\frac{\partial\pi_i(\vec{\varepsilon};\vec{v})}{\partial v_i}=0, i\in\{1,2,...,m\}$) yields the equilibrium strategies
\begin{align}
v^*_1& =\frac{v_2}{2}+\frac{y_1}{2c},\\		
v^*_i& =\frac{{cv_{i+1}}\delta_{i}+{cv_{i-1}}\delta_{i-1}+y_i}{2c[\delta_{i}+\delta_{i-1}]}, i\in\{2,..,m\},\\
v^*_m& =\frac{v_{m-1}}{2}+\frac{y_m}{2c}.
\end{align}


In the last stage, the SPs determine their privacy risk guarantees $\vec{\varepsilon}$ by considering equilibrium strategies in previous stages ($n_i$ and $ v^*_i \quad\forall i\in\{1,...,m\}$) as functions of $\vec{\varepsilon}$. Therefore, the optimal privacy risk strategy of $SP_i$ is determine by 
\begin{align}
\label{optme}
arg\max\limits_{\varepsilon_i}\pi_i(\vec{\varepsilon};\vec{v}), i\in\{1,...,m\}
\end{align}
while fixing all other players' strategies. For reasons of intractability (solving highly parameterized high order polynomial equations), a full characterization of privacy risk equilibria could not be achieved. Thus, we characterizes the SPNE numerically by using the iterated best response method. We consider a three-SP market and adopt the model parameters presented in Table ~\ref{simulationparameters}. Furthermore, we assume $t=0.7$ and $\bar{\varepsilon}=5$. The initial privacy risk of $SP_i$ is given by$\frac{i\bar{\varepsilon}}{i+1}$ for $i\in\{1,2,3\}$.  Although there exists an SPNE in the second stage of the sequential game for fixed privacy guarantees, the existence of an equilibrium in the first stage can not be guaranteed.  
\begin{figure}[h]%
	\centering
	\subfloat[Best response of each SP's privacy risk ($p_2=0.75$) ]{{\includegraphics[width=40mm]{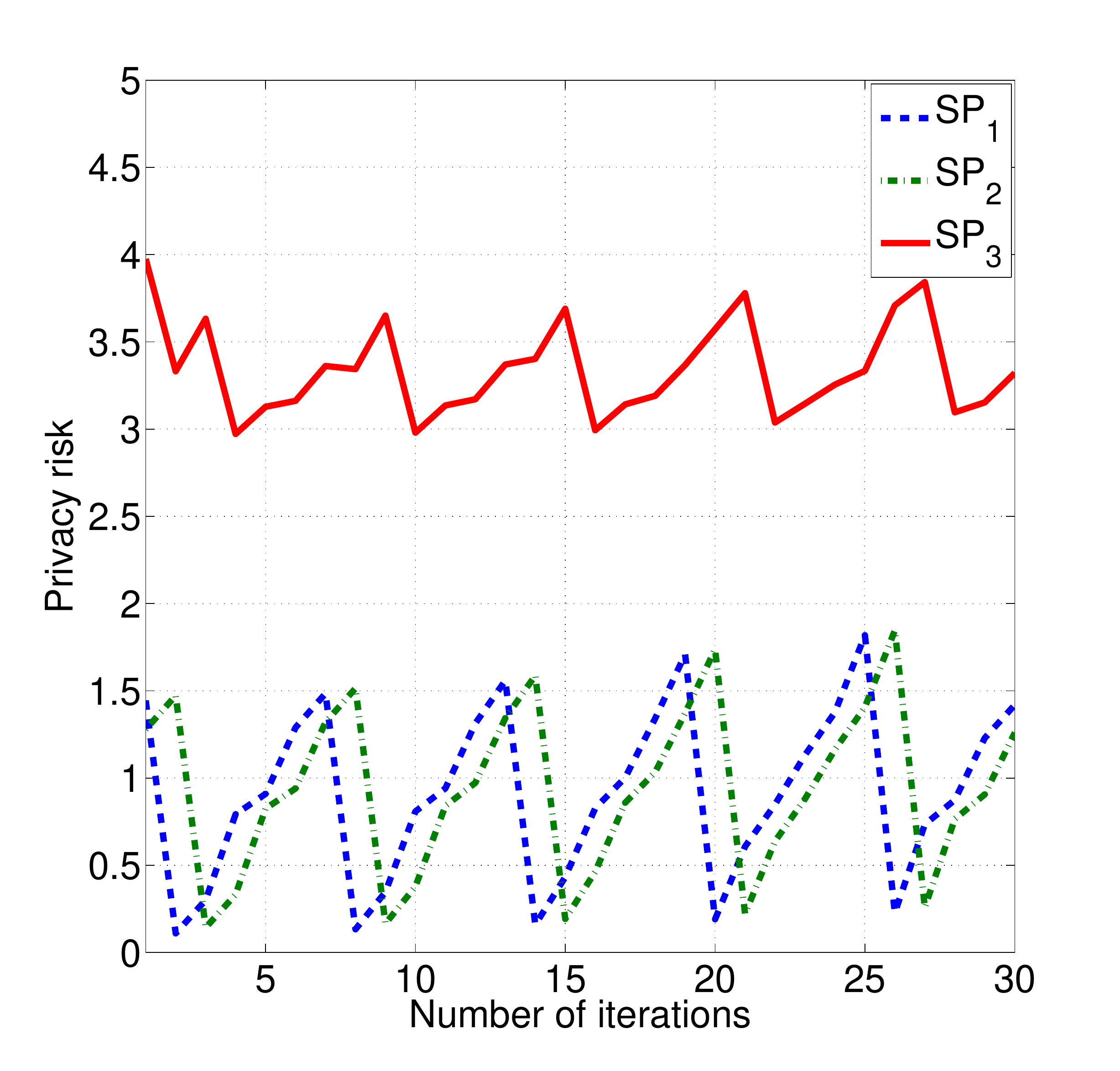} }\label{LargePcIter}}%
	\qquad
	\subfloat[Best response of each SP's privacy risk ($p_2=0.6$)]{{\includegraphics[width=40mm]{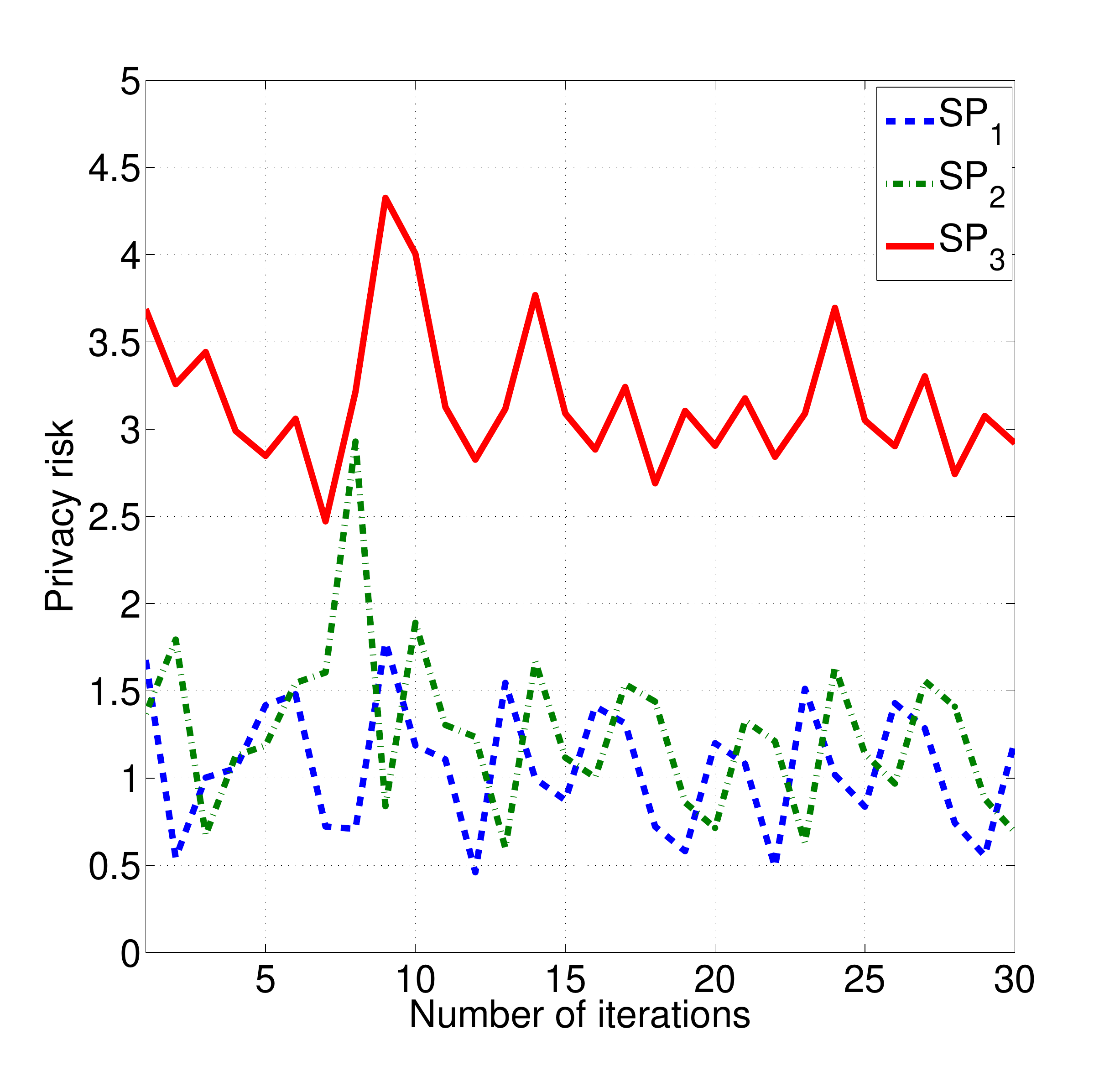} }\label{MidPcIter}}%
	\qquad
	\subfloat[Best response of each SP's privacy risk ($p_2=0.45$)]{{\includegraphics[width=40mm]{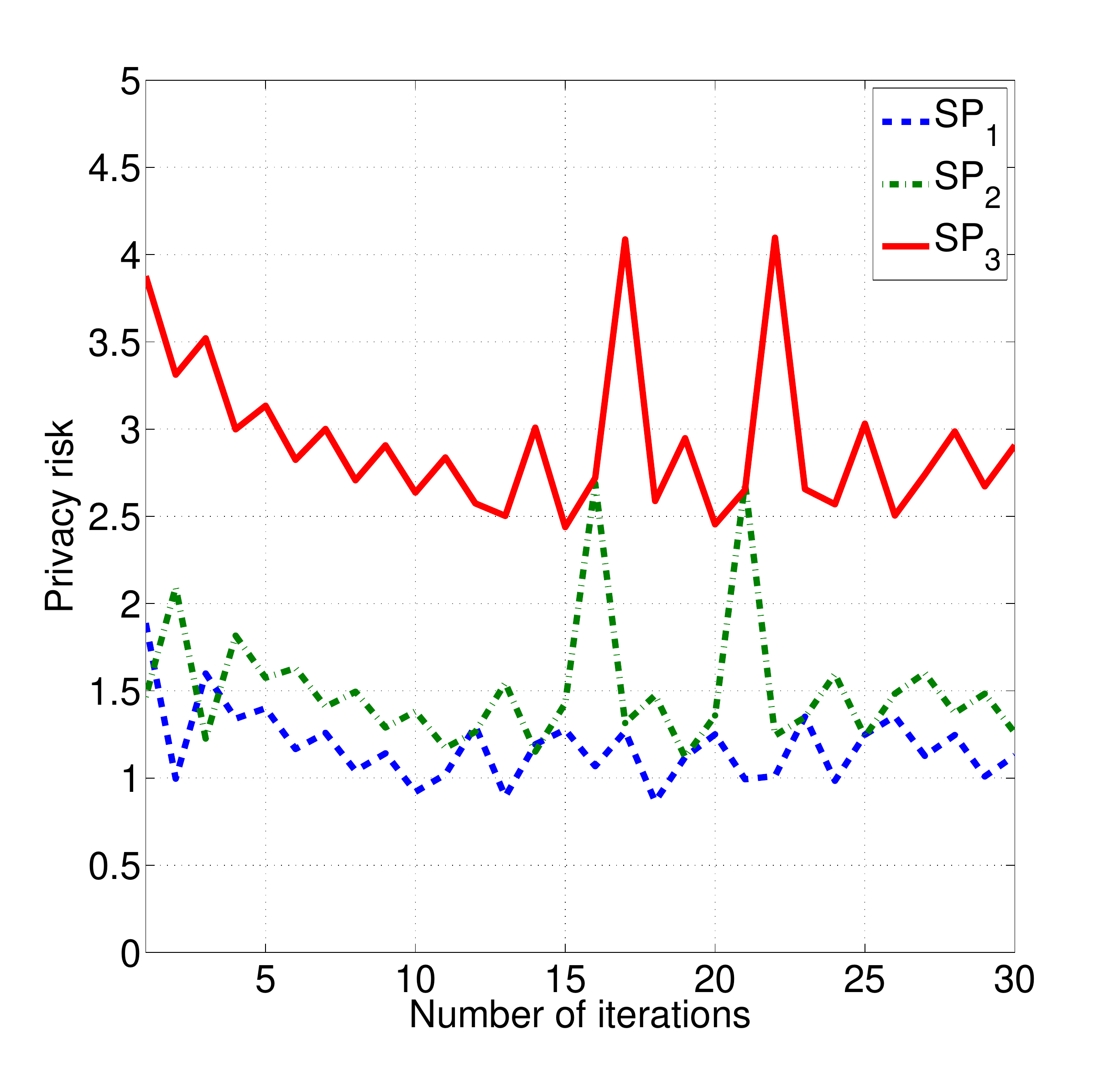} }\label{SmallPcIter}}%

	\caption{Best response of each SP's privacy risk for different values of $SP_2$'s revenue independent of using private data}%
	\label{BRPcIte}%
\end{figure}

The best response strategies of the SPs for different values of $SP_2$'s privacy independent revenue are plotted in Figure~\ref{BRPcIte}. It can be seen that 
the two SPs with lower privacy risks proceed to jump over each other in each round of best response iteration, attempting to lower its privacy risk to attract more consumers from its competitor. The SP with the highest revenue independent of using private data adopts a high privacy risk strategy to focus on consumers with high privacy risk tolerance and exploiting their private data extensively. Furthermore, we observe that when $p_2$ is large, $SP_3$'s privacy risk strategy is also higher on average. On the other hand, $SP_1$'s best response strategy is lower. The intuition behind is that a larger $p_2$ allows $SP_2$ to set a higher privacy risk to make more profit from using consumer data. This forces $SP_3$ to increase its privacy risk to differentiate itself from $SP_2$. On the other hand, a higher privacy risk of $SP_2$ will encourage $SP_1$ to lower its privacy risk to attract more consumers.

\section{Conclusions}
\label{Section:Conclusions}
Our work seeks to understand the effect of offering privacy- and QoS- differentiated online services on consumers with heterogeneous expressions of privacy sensitivity. We have quantified this effect as the fraction of consumers that prefer lower privacy risks with the accompanying lower QoS to the alternative of higher risks and higher QoS. We have presented an analysis built upon the classical Hotelling model to compute these fractions for both the two and multi SP problem. Analogous to the classical segmentation models, our problem also involves parameters that capture cost, revenue, and consumer valuation functions that are dependent and independent of the privacy risks. While such parametrized model can make the analysis challenging, our results for relatively simple yet meaningful functions such as linear cost models and uniform (as well as truncated Gaussian) distribution of consumer preferences suggests that SPs that have higher profits from untargeted services have an edge in the market. SPs competing on offering lower privacy risks have to offer better QoS or figure out other means of increasing untargeted revenue to gain market share. The market segmentation model assumes at least two or more SPs were able to overcome the barrier to entry and differentiate themselves. Thus, a related question we will address going forward is whether such barriers to entry are in fact surmountable when competitors use privacy as a differentiator. These analyses are crucial for developing better privacy policies to effectively enable safe and secure online commerce.

\section{Acknowledgements}
We wish to thank Professor Anand Sarwate at Rutgers University for many interesting discussions. This work is funded in part by the National Science Foundation under grant CCF-1350914.

%
%

\appendix
\section*{APPENDIX}
\section{Proof of Theorem~\ref{Theorem:SPNE2SP}}
\label{Proof:SPNE2SP}

Starting form the last stage in which consumers choose different SPs, we use backward induction to find the SPNE of the sequential game. In the last stage, each consumer located at $x\in[0,1]$ chooses an SP which maximize its utility function~\eqref{uservaluation}. By~\eqref{indifference} and the assumption that consumers' privacy risk tolerances are uniformly distributed, the indifference threshold $x_\tau$ is given by
\begin{align}
\label{indifferencelinear}
x_\tau=\frac{v_1-v_2+\frac{t(\varepsilon^2_2-\varepsilon_1^2)}{\bar{\varepsilon}}}{t(\varepsilon_2-\varepsilon_1)}=n_1(v_1;\varepsilon_1;v_2;\varepsilon_2).
\end{align} 

At the second stage, the optimal strategy of each SP is determined by the solution of \eqref{optD}. For fixed privacy risk guarantees $\varepsilon_2$ and $\varepsilon_1$,  the objective function of $SP_i, i\in\{1,2\}$ in this stage, i.e. $\pi_i(v_i;\varepsilon_i; v_{-i}; \varepsilon_{-i})$, is a concave function with respect to its own strategy $v_i$. Furthermore, the feasible set of $SP_i$'s strategy is a convex set ($v_i\in[0,+\infty]$). Thus, the non-cooperative subgame between $SP_2$ and $SP_1$ in this stage can be considered as a two-player concave game. By Theorem 1 and 2 in~\cite{rosen1965existence}, we can establish
\begin{proposition}
	For fixed privacy risk strategies, there exists a unique Nash equilibrium in the game between $SP_2$ and $SP_1$ at the second stage.	
\end{proposition}
To compute the equilibrium strategy of the second stage, we first substitute \eqref{clinear},  \eqref{vlinear}, and~\eqref{indifferencelinear} into \eqref{pii} and \eqref{pie}. Then, we apply the first order condition to SPs' profit functions and solve the simultaneous equations given by
\begin{align}
\label{derivativemsp}
\frac{\partial\pi_i(v_i;\varepsilon_i; v_{-i}; \varepsilon_{-i})}{\partial v_i}=0 \quad\forall i\in\{1,2\}.
\end{align}  
Solving the above simultaneous equations yields
\begin{align}	
	\label{vebr}
	& v_1=\frac{r\varepsilon_1+p_1}{2c}+\frac{v_2-\lambda\varepsilon_1-tx_2\varepsilon_2+tx_1\varepsilon_1}{2},\\	
	\label{vibr}
		& v_2=\frac{r\varepsilon_2+p_2}{2c}+\frac{v_1-\lambda\varepsilon_2-t(1-x_2)\varepsilon_2+t(1-x_1)\varepsilon_1}{2}.
\end{align}
For given privacy guarantees $\varepsilon_1$, and $\varepsilon_2$, solving the simultaneous linear equations above by substituting \eqref{vebr} into \eqref{vibr} yields the equilibrium strategies 
	\begin{align}	
	\label{ve} & v^*_1(\varepsilon_2,\varepsilon_1)=\frac{2(r\varepsilon_1+p_1)+r\varepsilon_2+p_2}{3c}+\frac{t(1+x_1)\varepsilon_1-\lambda(\varepsilon_2+2\varepsilon_1)-t(1+x_2)\varepsilon_2}{3},\\
	\label{vi}
	& v^*_2(\varepsilon_2,\varepsilon_1)=\frac{2(r\varepsilon_2+p_2)+r\varepsilon_1+p_1}{3c}+\frac{t(2-x_1)\varepsilon_1-\lambda(2\varepsilon_2+\varepsilon_1)-t(2-x_2)\varepsilon_2}{3}.
	\end{align}
	
	At the first stage, the SPs determine their optimal privacy risk by considering the QoS of each SP and the market segmentation computed in previous stages as functions of privacy risks offered by the SPs. By substituting \eqref{vi} and \eqref{ve} into \eqref{pii} and \eqref{pie}, the profit functions of the SPs can be written as
	\begin{align}
	\label{pibr}
	\pi_2& =\frac{c}{9t(\varepsilon_2-\varepsilon_1)}[\frac{p_2-p_1}{c}+(\frac{r}{c}-\lambda+t\frac{2\bar{\varepsilon}-\varepsilon_2-\varepsilon_1}{\bar{\varepsilon}})(\varepsilon_2-\varepsilon_1)]^2,\\
	\label{pebr}
	\pi_1& =\frac{c}{9t(\varepsilon_2-\varepsilon_1)}[-\frac{p_2-p_1}{c}+(-\frac{r}{c}+\lambda+t\frac{\bar{\varepsilon}+\varepsilon_2+\varepsilon_1}{\bar{\varepsilon}})(\varepsilon_2-\varepsilon_1)]^2.
	\end{align}
	Next, we apply the first order condition to SPs' profit functions to compute the equilibrium strategies. Taking the derivatives of $\pi_2$ and $\pi_1$ with respect to $\varepsilon_2$ and $\varepsilon_1$ and set both of their values to $0$ yields
	\begin{align}
	\frac{\partial\pi_2}{\partial\varepsilon_2}& =\frac{c[(\frac{r}{c}-\lambda+t\frac{2\bar{\varepsilon}-\varepsilon_2-\varepsilon_1}{\bar{\varepsilon}})(\varepsilon_2-\varepsilon_1)+\frac{p_2-p_1}{c}][(\frac{r}{c}-\lambda+t\frac{2\bar{\varepsilon}-3\varepsilon_2+\varepsilon_1}{\bar{\varepsilon}})(\varepsilon_2-\varepsilon_1)-\frac{p_2-p_1}{c}]}{9t(\varepsilon_2-\varepsilon_1)^2}=0,\\
	\frac{\partial\pi_1}{\partial\varepsilon_1}& =\frac{c[(-\frac{r}{c}+\lambda+t\frac{\bar{\varepsilon}+\varepsilon_2+\varepsilon_1}{\bar{\varepsilon}})(\varepsilon_2-\varepsilon_1)-\frac{p_2-p_1}{c}][(\frac{r}{c}-\lambda-t\frac{\bar{\varepsilon}-\varepsilon_2+3\varepsilon_1}{\bar{\varepsilon}})(\varepsilon_2-\varepsilon_1)-\frac{p_2-p_1}{c}]}{9t(\varepsilon_2-\varepsilon_1)^2}=0.
	\end{align}
	Solving the two simultaneous equations above yields
	\begin{align}
	\label{pdi1}
	&(\frac{r}{c}-\lambda+t\frac{2\bar{\varepsilon}-\varepsilon_2-\varepsilon_1}{\bar{\varepsilon}})(\varepsilon_2-\varepsilon_1)+\frac{p_2-p_1}{c}=0
	\end{align}
	or
	\begin{align}
	\label{pdi2}
	(\frac{r}{c}-\lambda+t\frac{2\bar{\varepsilon}-3\varepsilon_2+\varepsilon_1}{\bar{\varepsilon}})(\varepsilon_2-\varepsilon_1)-\frac{p_2-p_1}{c}=0
	\end{align}
	and 
	\begin{align}
	\label{pde1}
	(-\frac{r}{c}+\lambda+t\frac{\bar{\varepsilon}+\varepsilon_2+\varepsilon_1}{\bar{\varepsilon}})(\varepsilon_2-\varepsilon_1)-\frac{p_2-p_1}{c}=0
	\end{align}
	or
	\begin{align}
	\label{pde2}
	(\frac{r}{c}-\lambda-t\frac{\bar{\varepsilon}-\varepsilon_2+3\varepsilon_1}{\bar{\varepsilon}})(\varepsilon_2-\varepsilon_1)-\frac{p_2-p_1}{c}=0.
	\end{align}
	We note that the strategies given by \eqref{pdi1} and \eqref{pde1} result in $0$ profits in \eqref{pibr} and \eqref{pebr}. This indicates the privacy risk determined by \eqref{pdi1} and \eqref{pde1} are strictly dominated by the strategies given by the solution of \eqref{pdi2} and \eqref{pde2}. Solving \eqref{pdi2} and \eqref{pde2} yields the equilibrium privacy risk \eqref{optepsilinear} and 
	\begin{align}
	\label{optepse1}
	\varepsilon^*_1&=\frac{12\bar{\varepsilon}c\alpha-3ct\bar{\varepsilon}-16(p_2-p_1)}{24tc}.
	\end{align} 
	By subtracting  \eqref{optepse1} from \eqref{optepsilinear}, we have \eqref{optepselinear}. 
	Substitute the solution of $\varepsilon^*_2$ and $\varepsilon^*_1$ to \eqref{vi} and \eqref{ve}, we have \eqref{optvilinear} and 
	\begin{align}
	\label{optve1}
	v^*_1& =\frac{(2\alpha-t)c\alpha6\bar{\varepsilon}+(\alpha-3t)3ct\bar{\varepsilon}+(t-\alpha)16p_2+(2\alpha+t)8p_1}{24ct}.
	\end{align}
	Subtracting \eqref{optve1} from \eqref{optvilinear} yields \eqref{optvelinear}.
	
    Next, we prove the sufficient condition for the existence of the above SPNE. First of all, the model parameters must sustain a competitive market environment. Thus, in the equilibrium, each SP must have non-zero market share. This indicates the parameters must satisfy $0\le x^*_\tau=\frac{v^*_1-v^*_2+t(x^*_2\varepsilon^*_2-x^*_1\varepsilon^*_1)}{t(\varepsilon^*_2-\varepsilon^*_1)} \le 1$. Substitute \eqref{optepsilinear}, \eqref{optvilinear}, \eqref{optepselinear}, and \eqref{optvelinear} into the above inequality, we have \eqref{feasiblehotelling}. Furthermore, in the SPNE, the QoS of each SP must be non-negative (QoS feasibility) and the privacy risk guarantees must be bounded between $0$ and $\bar{\varepsilon}$ (privacy risk feasibility). By the model assumption in Section~\ref{Section:SPS}, we have $\varepsilon_1\le \varepsilon_2$. Thus, we only requires $\varepsilon_2\le\bar{\varepsilon}$ and $\varepsilon_1\ge 0$.  Substitute \eqref{optepsilinear} and \eqref{optepselinear} into the two inequalities above yields \eqref{feasibleeps}. Let  $x^*_i=\frac{\varepsilon^*_i}{\bar{\varepsilon}}, i\in\{A,B\}$ denote the normalized privacy risk of each SP in the SPNE. The equilibrium strategies must satisfy the complete market coverage condition given by $u_i(x)=v^*_i-t(x-x^*_i)\varepsilon^*_i\ge 0 \quad\forall x\in[0,1]$ for at least one $i\in\{A,B\}$.	
	
	Substituting \eqref{feasibleeps} into \eqref{optvelinear}, we have
	$v^*_2-v^*_1=\frac{3\bar{\varepsilon}}{4}\alpha-\frac{p_2-p_1}{3c}\ge\frac{3t\bar{\varepsilon}}{16}+\frac{2(p_2-p_1)}{3c}>0$, thus we only need $v_1\ge0$ for QoS feasibility. Furthermore, the Hotelling model feasibility condition implies $v^*_1-tx^*_1\varepsilon^*_1\ge v^*_2-tx^*_2\varepsilon^*_2$. Since $u_i(x)$ is an increasing function of $x$, complete market coverage condition can be simplified to $u_1(0)\ge0$. As a result, the QoS feasibility condition and the complete market coverage condition can be simplified to $v^*_1-tx^*_1\varepsilon^*_1\ge 0$. Therefore, the sufficient condition for the existence of SPNE is given by:
	\begin{enumerate}
	\item $0 \le\frac{v^*_1-v^*_2+t(x^*_2\varepsilon_2-x^*_1\varepsilon_1)}{t(\varepsilon^*_2-\varepsilon^*_1)}\le 1$,
	\item $0 \le \varepsilon^*_1, \varepsilon^*_2 \le\bar{\varepsilon}$,
	\item $v^*_1-tx^*_1\varepsilon^*_1\ge 0$.			
	\end{enumerate}	
	Solving the above three inequalities yield \eqref{feasiblehotelling}, \eqref{feasibleeps}, and \eqref{mktcoverage}.
	The equilibrium market share and profits of the SPs are obtained by substituting \eqref{optepsilinear}, \eqref{optvilinear}, \eqref{optepselinear}, and \eqref{optvelinear} into \eqref{indifference}, \eqref{pie}, and \eqref{pii}.

%



%


\bibliographystyle{IEEEtran}
\bibliography{IEEEabrv,PaperRefs}

\begin{thebibliography}{10}
\providecommand{\url}[1]{#1}
\csname url@samestyle\endcsname
\providecommand{\newblock}{\relax}
\providecommand{\bibinfo}[2]{#2}
\providecommand{\BIBentrySTDinterwordspacing}{\spaceskip=0pt\relax}
\providecommand{\BIBentryALTinterwordstretchfactor}{4}
\providecommand{\BIBentryALTinterwordspacing}{\spaceskip=\fontdimen2\font plus
\BIBentryALTinterwordstretchfactor\fontdimen3\font minus
  \fontdimen4\font\relax}
\providecommand{\BIBforeignlanguage}[2]{{%
\expandafter\ifx\csname l@#1\endcsname\relax
\typeout{** WARNING: IEEEtran.bst: No hyphenation pattern has been}%
\typeout{** loaded for the language `#1'. Using the pattern for}%
\typeout{** the default language instead.}%
\else
\language=\csname l@#1\endcsname
\fi
#2}}
\providecommand{\BIBdecl}{\relax}
\BIBdecl

\bibitem{Erlingsson2014}
\BIBentryALTinterwordspacing
\'{U}lfar Erlingsson, V.~Pihur, and A.~Korolova, ``{RAPPOR}: Randomized
  aggregatable privacy-preserving ordinal response,'' in \emph{Proceedings of
  the 2014 ACM SIGSAC Conference on Computer and Communications Security (CCS
  '14)}, 2014, pp. 1054--1067. [Online]. Available:
  \url{http://dx.doi.org/10.1145/2660267.2660348}
\BIBentrySTDinterwordspacing

\bibitem{AppleSupport2016}
``About privacy and location services in ios 8 and later,''
  https://support.apple.com/en-is/HT203033, Sep 2016.

\bibitem{AcquisiTW:16econ}
\BIBentryALTinterwordspacing
A.~Acquisti, C.~Taylor, and L.~Wagman, ``The economics of privacy,''
  \emph{Journal of Economic Literature}, vol.~54, no.~2, pp. 442--492, jun
  2016. [Online]. Available: \url{http://dx.doi.org/10.1257/jel.54.2.442}
\BIBentrySTDinterwordspacing

\bibitem{shaffer1995competitive}
G.~Shaffer and Z.~J. Zhang, ``Competitive coupon targeting,'' \emph{Marketing
  Science}, vol.~14, no.~4, pp. 395--416, 1995.

\bibitem{chen2002research}
Y.~Chen and G.~Iyer, ``Research note consumer addressability and customized
  pricing,'' \emph{Marketing Science}, vol.~21, no.~2, pp. 197--208, 2002.

\bibitem{tang2008gaining}
Z.~Tang, Y.~Hu, and M.~D. Smith, ``Gaining trust through online privacy
  protection: Self-regulation, mandatory standards, or caveat emptor,''
  \emph{Journal of Management Information Systems}, vol.~24, no.~4, pp.
  153--173, 2008.

\bibitem{campbell2015privacy}
J.~Campbell, A.~Goldfarb, and C.~Tucker, ``Privacy regulation and market
  structure,'' \emph{Journal of Economics \& Management Strategy}, vol.~24,
  no.~1, pp. 47--73, 2015.

\bibitem{conitzer2012hide}
V.~Conitzer, C.~R. Taylor, and L.~Wagman, ``Hide and seek: Costly consumer
  privacy in a market with repeat purchases,'' \emph{Marketing Science},
  vol.~31, no.~2, pp. 277--292, 2012.

\bibitem{chen2001individual}
Y.~Chen, C.~Narasimhan, and Z.~J. Zhang, ``Individual marketing with imperfect
  targetability,'' \emph{Marketing Science}, vol.~20, no.~1, pp. 23--41, 2001.

\bibitem{jentzsch2012study}
N.~Jentzsch, S.~Preibusch, and A.~Harasser, ``Study on monetising privacy: An
  economic model for pricing personal information,'' \emph{ENISA, Feb}, 2012.

\bibitem{lee2011managing}
D.-J. Lee, J.-H. Ahn, and Y.~Bang, ``Managing consumer privacy concerns in
  personalization: a strategic analysis of privacy protection,''
  \emph{Management Information Systems Quarterly}, vol.~35, no.~2, pp.
  423--444, 2011.

\bibitem{chellappa2010mechanism}
\BIBentryALTinterwordspacing
R.~K. Chellappa and S.~Shivendu, ``Mechanism design for ``free'' but ``no free
  disposal'' services: The economics of personalization under privacy
  concerns,'' \emph{Management Science}, vol.~56, no.~10, pp. 1766--1780, 2010.
  [Online]. Available: \url{http://dx.doi.org/10.1287/mnsc.1100.1210}
\BIBentrySTDinterwordspacing

\bibitem{DattaTD:pets2015}
\BIBentryALTinterwordspacing
A.~Datta, M.~C. Tschantz, and A.~Datta, ``Automated experiments on ad privacy
  settings: A tale of opacity, choice, and discrimination,'' in
  \emph{Proceedings on Privacy Enhancing Technologies}, vol. 2015, no.~1, apr
  2015, pp. 92--112. [Online]. Available:
  \url{http://dx.doi.org/10.1515/popets-2015-0007}
\BIBentrySTDinterwordspacing

\bibitem{perloff2016microeconomics}
J.~M. Perloff, \emph{Microeconomics: theory and applications with
  calculus}.\hskip 1em plus 0.5em minus 0.4em\relax Pearson, 2016.

\bibitem{hotelling1990stability}
H.~Hotelling, ``Stability in competition,'' in \emph{The Collected Economics
  Articles of Harold Hotelling}.\hskip 1em plus 0.5em minus 0.4em\relax
  Springer, 1990, pp. 50--63.

\bibitem{wauthy1996quality}
X.~Wauthy, ``Quality choice in models of vertical differentiation,'' \emph{The
  Journal of Industrial Economics}, pp. 345--353, 1996.

\bibitem{brenner2005hotelling}
S.~Brenner, ``Hotelling games with three, four, and more players,''
  \emph{Journal of Regional Science}, vol.~45, no.~4, pp. 851--864, 2005.

\bibitem{fudenberg1991game}
D.~Fudenberg and J.~Tirole, \emph{Game theory}.\hskip 1em plus 0.5em minus
  0.4em\relax MIT press, MA, 1991.

\bibitem{osborne1994course}
M.~J. Osborne and A.~Rubinstein, \emph{A course in game theory}.\hskip 1em plus
  0.5em minus 0.4em\relax MIT press, 1994.

\bibitem{rosen1965existence}
J.~B. Rosen, ``Existence and uniqueness of equilibrium points for concave
  n-person games,'' \emph{Econometrica: Journal of the Econometric Society},
  pp. 520--534, 1965.

\end{thebibliography}
\end{document}